\pdfoutput=1
\documentclass[11pt,twoside,a4paper,cmspaper,final,collab]{cms-tdr}
\def\svnVersion{7b3171b}\def\svnDate{2025/01/08}\def\cmsCernNoTag{CERN-EP-2023-219}\def\cmsCernDate{\today}\def\cmsMessage{Published in the Journal of High Energy Physics as \href{http://dx.doi.org/10.1007/JHEP01(2025)011}{\doi{10.1007/JHEP01(2025)011}.}}
\begin{document}\cmsNoteHeader{SMP-21-009}

\cmsNoteHeader{SMP-21-009}
\title{Measurement of the double-differential inclusive jet cross section in proton-proton collisions at \texorpdfstring{$\sqrt{s}  =  5.02\TeV$}{sqrt(s) = 5.02 TeV}}

\date{\today}

\abstract{
The inclusive jet cross section is measured as a function of jet transverse momentum \pt and rapidity $y$. The measurement is performed using proton-proton collision data  at $\sqrt{s} = 5.02\TeV$, recorded by the CMS experiment at the LHC, corresponding to an integrated luminosity of 27.4\pbinv. The jets are reconstructed with the anti-\kt algorithm using a distance parameter of $R=0.4$, within the rapidity interval $\abs{y} < 2$, and across the kinematic range $0.06 < \pt < 1\TeV$. The jet cross section is unfolded from detector to particle level using the determined jet response and resolution. The results are compared to predictions of perturbative quantum chromodynamics, calculated at both next-to-leading order and next-to-next-to-leading order. The predictions are corrected for nonperturbative effects, and presented for a variety of parton distribution functions and  choices of the renormalization/factorization scales and the strong coupling \alpS.
}

\hypersetup{%
pdfauthor={CMS Collaboration},%
pdftitle={Measurement of the double-differential inclusive jet cross section in proton-proton collisions at sqrt(s) = 5.02 TeV},%
pdfsubject={CMS, jet cross section, 5 TeV},%
pdfkeywords={CMS, jets, QCD, 5 TeV}} 

\maketitle 
\section{Introduction}
\label{sec:intro}

One of the most fundamental standard model (SM) measurements is that of the production cross section of jets in 
hadron collisions. When performed double-differentially, as a function of both transverse momentum (\pt) and 
rapidity ($y$) of the produced jets, 
it provides an opportunity to thoroughly study the jet kinematics. In the inclusive jet production,
almost all of the jets are initiated from pure quantum chromodynamics (QCD) processes, with negligible background
coming from intermediate electroweak processes with jet final states. This allows for a direct comparison of experimental
results with perturbative QCD (pQCD) calculations corrected for the nonperturbative (NP) effects of hadronization 
and multiparton interactions. Since the final measured hadronic cross section may be described as a convolution 
of the partonic cross section with parton distribution functions (PDFs) of the initial-state hadrons, the measurement offers constraints on PDFs and the strong coupling (\alpS), as well as a better understanding of the factorization 
and renormalization scales. It is also important to perform the measurement at different values of the center-of-mass energy 
($\sqrt{s}$) of the colliding hadrons, because this affects the energy scales and strength of the interactions.
In this paper, we present the measurement of the double-differential cross section of inclusive jet production 
in proton-proton (pp) collisions at $\sqrt{s}=5.02$\TeV with an integrated luminosity of 27.4\pbinv.
The measurement is performed with data collected in 2015 by the 
CMS experiment during a special lower-energy run of the Large Hadron Collider (LHC) at CERN.

Measurements of the inclusive jet cross section have been performed: in pp collisions at 
the CERN Intersecting Storage Rings (ISR) at $\sqrt{s}$ equal to 45\GeV~\cite{AFS_45}, and 63\GeV~\cite{AFS_63_1}; in
proton-antiproton collisions at the CERN Super Proton Antiproton Synchrotron (S$\mathrm{p}\bar{\mathrm{p}}$S) 
at 540--546\GeV\cite{UA1_540,UA2_546},
630\GeV~\cite{UA1_546_630, UA2_630}, and 200--900\GeV~\cite{UA1_200_900}; in proton-antiproton collisions at the Fermilab Tevatron at 546\GeV~\cite{CDF_546}, 630\GeV~\cite{CDF_630,D0_630}, 1.8\TeV~\cite{CDF_1800, D0_1800}, and 1.96\TeV~\cite{CDF_1960,CDF_1960_2, CDF_1960_3, D0_1960, D0_1960_2}; and in pp collisions at the LHC at 2.76\TeV~\cite{ALICE_276, ATLAS_276, CMS_276}, 
7\TeV~\cite{ATLAS_7,CMS_7,ATLAS_7_2,CMS_7_2,CMS_7_3,ATLAS_7_3,CMS_7_4}, 
8\TeV~\cite{CMS_8, ATLAS_8}, and 13\TeV~\cite{CMS_13, ATLAS_13, CMS_13_2, CMS_13_3, CMS_13_3b}.
All these measurements led to a better understanding of pQCD within the constraints of NP QCD effects.
At the same time, they provided accurate estimations of QCD-initiated backgrounds in both SM analyses and direct searches for new physics, 
and offered the possibility of indirect discoveries.

The presented cross section measurement has the added value of providing a reference for the respective 5.02\TeV heavy-ion jet
analyses probing the quark-gluon plasma~\cite{1999401, mT_strange_PbPb_158GeV_WA97, 2017193, JER_SF, HIN5TeV}. In such plasma conditions,
because of jet quenching~\cite{d'Enterria:2009am}, the resulting jets demonstrate differences in the 
parton showering~\cite{PhysRevLett.120.142302}, fragmentation~\cite{PhysRevLett.121.242301}, and momentum imbalance in back-to-back dijets~\cite{PhysRevLett.105.252303, 2012176}, compared with the respective jets produced in pp collisions.

The paper is structured as follows. In Section~2, we present the cross section measurement strategy and in Section~3 we give a brief description of the CMS experiment. The jet reconstruction, identification, and resolution are described in Section~4, and in Section~5 we present the online data collection and trigger efficiencies. The simulation and detector-level spectra are discussed in Sections 6 and 7, respectively. In Section~8, the pQCD prediction and the estimation of the NP effects are given. In Section~9, we discuss the procedure of unfolding the detector-level spectra to particle-level. The experimental and theoretical systematic uncertainties are detailed in Section~10. The final experimental results and the comparisons with theoretical predictions are shown in Section~11, and a summary is given in Section~12.  Tabulated results are provided in the HEPData record for this analysis~\cite{hepdata}.

\section{The double-differential inclusive jet cross section}

The inclusive jet cross section, differential in jet \pt and $y$,
is defined as :
\begin{equation}
  \frac{\rd^2\sigma}{\rd\pt \rd{y}} = \frac{N_\text{jets}}{\epsilon \lumi \Delta \pt \Delta y},
\end{equation}
where $N_\text{jets}$ is the number of jets observed within the 
bins of \pt and $y$ (that have widths $\Delta \pt$ 
and $\Delta y$, respectively), \lumi is the integrated luminosity of the data set, and $\epsilon$ is the product of the
event selection, jet selection, and trigger efficiencies.
In this measurement, the bin width $\Delta \pt$ varies with \pt according 
to the momentum resolution of the detector and the size of the available data sample.
The range of $y$ in this analysis is $\abs{y} < 2$, and the cross section is measured in four bins of $\abs{y}$, each with width of $\Delta\abs{y} = 0.5$
(thus $\Delta y = 1$).

Equation (1) can be applied in a straightforward manner to the detector-level jet cross section,
where jets are reconstructed from detected particles and are counted in bins of \pt vs. $\abs{y}$. On the other hand,
the desirable quantity to be measured is the particle-level jet cross section, where jets are reconstructed 
from stable particles (with mean path length $c\tau > 10\mm$), not affected by detector effects.
This is achieved by correcting the detector-level jet count in each \pt bin 
for bin migrations due to the limited detector momentum resolution
(migrations in $\abs{y}$ are negligible).
The transition from detector-level to particle-level
jet cross section is achieved with an unfolding procedure.
The particle-level cross section measurement is the final result, to be compared
with theoretical predictions.

\section{The CMS experiment}

The CMS experiment uses a right-handed coordinate system, with the origin at the nominal
interaction point, the $x$ axis pointing to the center of the LHC ring
and the $z$ axis along the counterclockwise-beam direction as viewed from above. The polar
angle $\theta$ is measured from the positive z axis and the azimuthal angle $\phi$ 
is measured on the $x$-$y$ plane. The pseudorapidity $\eta$ is defined as $-\ln(\tan(\theta/2))$.

The central feature of the CMS apparatus is a superconducting solenoid of 6\unit{m} internal diameter, providing a
magnetic field of 3.8\unit{T}. Within the magnetic volume are a
silicon pixel and strip tracker, a lead tungstate crystal
electromagnetic calorimeter (ECAL), and a brass and
scintillator hadron calorimeter (HCAL), each composed of a barrel
and two endcap sections. Forward calorimeters extend the
$\eta$ coverage provided by the barrel and endcap detectors. 
Muons are detected in gas-ionization chambers embedded 
in the steel flux-return yoke outside the
solenoid.

The ECAL consists of 75848 lead tungstate crystals, which provide coverage in $\abs{\eta} <
1.48$ in the barrel region and $1.48 < \abs{\eta} < 3.00$ in two endcap regions. Preshower detectors
consisting of two planes of silicon sensors interleaved with a total of 3 radiation lengths of lead are located
in front of each ECAL endcap detector.

In the region $\abs{\eta} < 1.74$, the HCAL cells have widths of 0.087 in both $\abs{\eta}$ and $\phi$ in radians. 
In the $\eta$-$\phi$ plane, and for $\abs{\eta} < 1.48$, the HCAL cells map onto 
$5{\times}5$ arrays of ECAL crystals to form calorimeter towers projecting radially outwards from 
close to the nominal interaction point. For $\abs{\eta} > 1.74$, the coverage of the towers increases 
progressively to a maximum of 0.174 in $\Delta \eta$ and $\Delta \phi$ in radians. Within each tower, the energy 
deposits in ECAL and HCAL cells are summed to define the calorimeter tower energies, which are subsequently 
used to provide the energies and directions of hadronic jets. When combining information from the entire detector, 
the jet energy resolution amounts typically to 15--20\% at 30\GeV, 10\% at 100\GeV, and 5\% at 1\TeV~\cite{Khachatryan:2016kdb}.

Events that are considered potentially
interesting are selected online by a two-tiered trigger system~\cite{Khachatryan:2016bia}.
The first level (L1), composed of custom hardware processors, uses information from the calorimeters and muon detectors to select events at a rate of around 100\unit{kHz} within a fixed latency of about 4\mus~\cite{Sirunyan:2020zal}. The second level, known as the high-level trigger (HLT)~\cite{hlt}, consists of a farm of processors running a version of the full event reconstruction software optimized for fast processing, and reduces the event rate to around 1\unit{kHz} before data storage.
A more detailed description of the CMS detector can be found in Ref.~\cite{cms}.

\section{Jet reconstruction, identification, and resolution\label{secJetId}}

Final-state particles are reconstructed using the particle-flow (PF) algorithm~\cite{pf}, which combines signals from all CMS subdetectors.
The energy of the particles is obtained as follows: 
\begin{itemize}
\item for photons, from the ECAL measurement;
\item for electrons, from a combination of the electron momentum at the primary interaction vertex as determined by the tracker, the energy of the corresponding ECAL cluster, and the energy sum of all bremsstrahlung photons compatible in space and time with originating from the electron track;
\item for muons, from the curvature of the corresponding track measured in both the tracker and the muon system;
\item for charged hadrons, from a combination of their momentum measured in the tracker and the matching ECAL and HCAL energy deposits. This energy is corrected for the response function of the calorimeters to hadronic showers;
\item for neutral hadrons, from the corresponding corrected ECAL and HCAL energies.
\end{itemize}

These particles are used as constituents in the anti-\kt jet reconstruction
algorithm~\cite{antiKt}, implemented in \FASTJET~\cite{fastJet}, with a distance parameter $R=0.4$. 
Charged hadrons originating from vertices other than the primary vertex (PV) are removed to 
reduce effects from multiple interactions per proton bunch crossing of the colliding beams (pileup).
The PV is required to be reconstructed from at least five charged-particle tracks
and must satisfy a set of quality requirements, including
$\abs{z_\mathrm{PV}} < 24\cm$ and $\rho_\mathrm{PV} < 2\cm$, where $z_\mathrm{PV}$ and $\rho_\mathrm{PV}$ 
are the longitudinal and transverse distances of the PV
from the nominal interaction point in the CMS detector. 
In the case that multiple vertices pass these requirements, 
the PV is taken to be the vertex corresponding to the hardest scattering in the event, evaluated using tracking information alone, as described in Section 9.4.1 of Ref.~\cite{CMS-TDR-15-02}.

In order to reduce the probability of misidentification, the jets used in this analysis are required to
satisfy the following quality requirements. Jets are required to have at least two PF-constituents, with 
at least one being charged. The fractions of neutral hadron energy, neutral electromagnetic energy, 
and charged electromagnetic energy of the jets each must not exceed 99\% of the jet's total energy. 
The muon energy fraction is required to be less than 80\% of this total energy and the charged-hadron energy must be nonzero.

The jet momenta used in the above fractions are the raw jet momenta, determined as the vectorial sum of all constituents' particle momenta. Because of calorimeter nonlinearities, pileup effects, and excess energy from the underlying event, 
the raw measured energy of the jets is corrected in a factorized approach using dedicated data samples and Monte Carlo (MC) simulations.
There are two levels of corrections for jets both in MC simulation and data:
a correction for pileup and electronic noise; and a correction for the response of the detector as a function of jet \pt and $\eta$.
An additional residual correction is applied only to jets in data, to account for differences between data and MC~\cite{JEC}.
These jet energy scale corrections (JES) are derived from simulation studies so that the average measured energy of jets becomes identical to that 
of particle-level jets. In situ measurements of the momentum balance in dijet, photon+jet, \Z+jet, and multijet events are 
used to determine any residual differences between the jet energy scale in data and in simulation, and appropriate 
corrections are made~\cite{Khachatryan:2016kdb}. The corrected energy is used in the measured and simulated jet \pt distributions presented.

An important aspect of this analysis is the determination of the jet energy resolution (JER) as a
function of jet \pt and $y$. The JER provides the correlation between particle-level jet energy
and the corresponding detector-level one, needed for unfolding the results to particle level.
The JER is determined by first creating the ratios 
of detector-level jet \pt over the corresponding particle-level value, $\pt^{\text{det}}/\pt^{\text{part}}$, 
using simulation, in bins of $\pt^{\text{part}}$ and rapidity.
The cores of these distributions are fitted with Gaussian distributions with varied normalization,
mean ($\mu_\mathrm{G}$) and standard-deviation ($\sigma_\mathrm{G}$) values.
These JER Gaussian functions are used for spreading the jet \pt generated via pseudo-experiments
during the construction of the response matrices. In order to achieve a smoother unfolding, the 
$\mu_\mathrm{G}$ and $\sigma_\mathrm{G}$ distributions as a function of \pt are fitted with continuous curves,
in each rapidity bin separately, so that a more accurate spreading is achieved for each pseudo-experiment-generated \pt.
A more complicated approach, which uses double-sided Crystal Ball functions with the addition of extra tails
in order to describe the detector effects,
was also implemented, with consistent unfolded results. The simpler Gaussian method is preferred since 
it provides smoother unfolded results, because of the fits of the $\mu_\mathrm{G}$ and $\sigma_\mathrm{G}$ \pt distributions.
The extracted JER values ($\sigma_\mathrm{G}$) are corrected with a scale factor of $1.1 \pm 0.1$, in order to account for 
the differences in jet \pt resolution between data and MC simulation, as was done in Ref.~\cite{JER_SF}.
The 10\% systematic uncertainty in JER is derived from the variation of the spreading of jet \pt in the MC dijet 
spectrum so that the latter matches the observed one in the data. The estimation of the resolution in the data
is obtained by comparing the \pt of the leading two jets in events with low extra jet activity.

\section{Data collection and trigger efficiencies\label{secData}}

We analyze data collected during a special LHC pp run at $\sqrt{s}=5.02\TeV$,
characterized by low instantaneous luminosity
that resulted in an average 1.1 collisions per bunch crossing with a reconstructed primary vertex. The corresponding integrated luminosity of the data set is 27.4\pbinv.

{\tolerance=800
The HLT selection consists of three separate single-jet trigger paths, 
designated here as ak4PFjet\{40,60,80\}.
The three triggers each require at least one jet in the event 
(reconstructed at HLT level) with transverse momentum 
$\pt^\mathrm{HLT}$ larger than 40, 60, and 80\GeV, respectively.
The respective thresholds at L1 level were 28, 40, and 48\GeV.
The 40 and 60\GeV triggers are prescaled, \ie, their data output rate 
is algorithmically reduced to accommodate the allocated bandwidth.
This means that the jet yields from those triggers 
have to be scaled up to account for the full integrated luminosity. In order to remove trigger-efficiency dependence on 
the measured jet rates, we utilize 
triggers for jet \pt above a threshold where they are fully efficient (efficiency $>$99.9\%).
We measure each trigger efficiency as a function of the triggered jet \pt
using a tag-and-probe method applied on the same mutlijet data.
In this method, we tag a jet that passed the trigger in question, we remove it, and we use the remaining (probe) jets to form a jet-\pt-dependent ratio (trigger efficiency) of number of jets that pass the trigger divided by the number of jets that pass some looser requirements (\eg, L1 trigger or a lower-\pt trigger).
We conclude that the 
ak4PFjet\{40,60,80\} triggers are fully efficient for jet \pt above 55, 75, and 105\GeV, respectively.
The proper merging of the data collected from these three triggers gives the 
detector-level jet spectra presented in Section~\ref{secDetSpectra}.
\par}

We select for further processing events that are not associated with 
noise, which could come from either the calorimeters or beam halo effects. 
We achieve this by requiring that the ratio of the momentum imbalance in the transverse plane to the 
scalar sum of the PF constituents' transverse momentum is less than 0.3, as required also in Refs.~\cite{CMS_276, CMS_7_2, CMS_8}. This selection 
requirement also suppresses jets associated with the production and decay of electroweak gauge bosons and top quarks.
The final data set includes events with the presence of a PV and with at least one jet passing the 
identification requirements presented in Section~\ref{secJetId}.

\section{Monte Carlo simulation\label{secMC}}

Monte Carlo techniques are used to simulate events in the CMS detector, as well as for verifications and corrections.
\PYTHIA8~\cite{pythia} generates jets with $2\to2$ leading-order (LO) Feynman diagrams, \pt-ordered dipole showering of outgoing partons,
and subsequent hadronization using the Lund string model~\cite{lund}. The NNPDF2.3~\cite{nnpdf23} LO PDF set is used along with the 
Monash~\cite{monash}-based CUETP8M1 underlying event tune~\cite{cuetp8m1} for \PYTHIA8, which uses CMS minimum bias data to describe the underlying event.
\HERWIG7~\cite{herwig} is used at LO $2\to2$ mode. 
The NNPDF3.1~\cite{nnpdf31} next-to-next-to-LO (NNLO) PDF set 
with $\alpS(m_{\Z})=0.118$ is used (where $m_{\Z}$ is the pole mass of the \Z boson), and the
NNPDF3.1 LO PDF set is used for the determination 
of multiparton interactions.
Angular-ordered showers are hadronized based 
on the cluster model~\cite{cluster}. 
The CH2 tune~\cite{ch2} is used for the description of the underlying event.
The events produced by the above generators pass through a \GEANTfour~\cite{geant4}-based CMS detector simulation giving signals that are used for the final reconstructed detector-level objects.

The \PYTHIA8 simulation, version 212, is used for an initial comparison
of the spectra measured in data and simulation at the detector level (Section~\ref{secDetSpectra}),
and for the determination of the JER (Section~\ref{secJetId}).
The \PYTHIA8 simulation, version 219, is used for the estimation of the NP corrections to the pQCD calculation
(Section~\ref{secQCD}). The \HERWIG7 simulation, version 2.0, is used for an alternate
estimation of the NP effects.

\section{Detector-level spectra\label{secDetSpectra}}

The inclusive detector-level jet \pt spectra, constructed from data collected with the three single-jet HLT paths, 
have to be merged so that the resulting spectra correspond to the true respective jet \pt
distributions, without any effects from trigger inefficiencies or trigger prescales, and without any double-counting
of events or jets.

\begin{figure}[th!]
\center
 \includegraphics[width=0.85\textwidth]{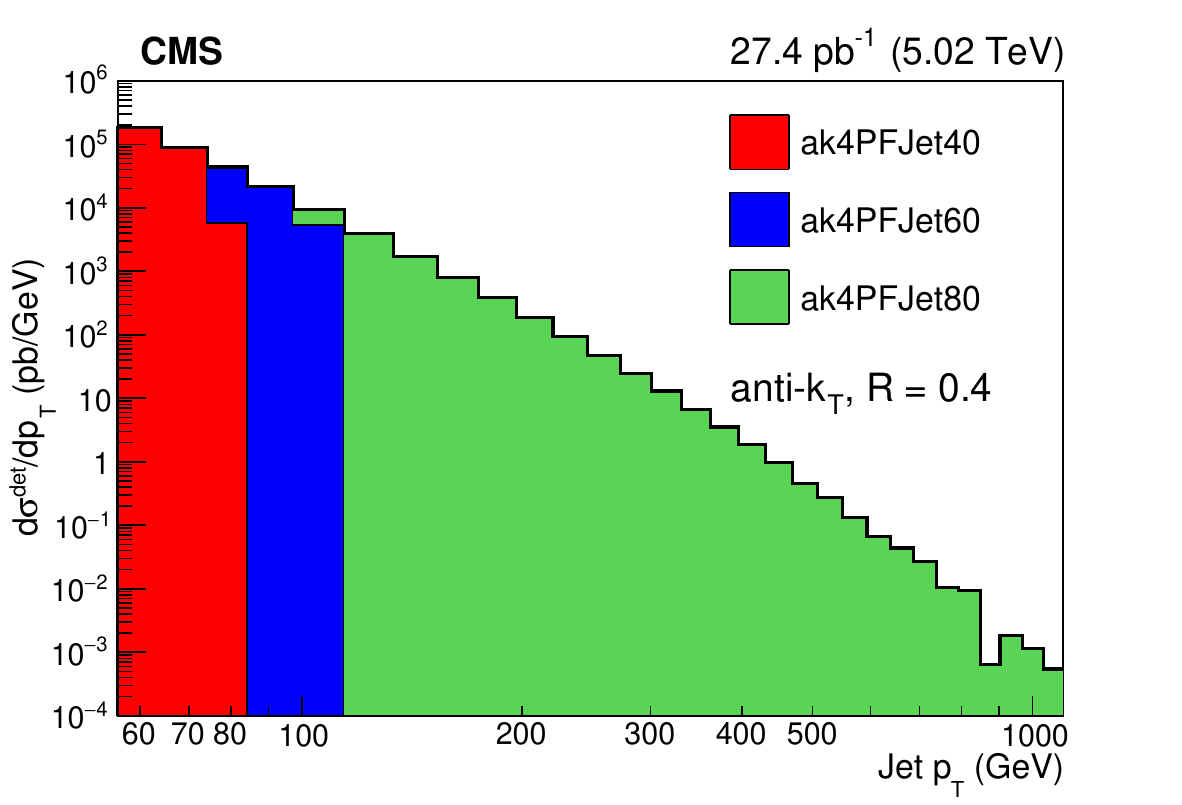}
\caption{
Detector-level cross section obtained after merging the contributions from the three triggers, normalized to their respective integrated luminosities.\label{merging}}
\end{figure}

In this analysis, we create three separate detector-level jet spectra, using data from each trigger
as a separate experiment with its own integrated luminosity, which is determined by integrating the 
instantaneous luminosity with the corresponding event-by-event prescales. In the separate spectra from the
ak4PFjet\{40,60,80\} triggers, jets with \pt above 55, 75, 105\GeV, respectively, are included. 
Subsequently, the three spectra are merged, by keeping only the jets from the three spectra 
that belong in the three full-efficiency \pt windows, $56 < \pt < 75\GeV$, $75 < \pt < 105\GeV$, and $\pt > 105\GeV$, 
respectively. The spectra are divided by their luminosities, to provide the detector-level cross sections, as shown below:
\begin{equation}
  \frac{\rd^2\sigma^\text{det}}{\rd\pt \rd{y}} = \frac{1}{\epsilon \Delta \pt \Delta y}
  \begin{cases}
   \frac{N_\text{jets}^\mathrm{jet40}}{\lumi^\mathrm{jet40}} & \text{if } 56 < \pt < 75\GeV\\
                \frac{N_\text{jets}^\mathrm{jet60}}{\lumi^\mathrm{jet60}} & \text{if } 75 < \pt < 105\GeV \\
	  \frac{N_\text{jets}^\mathrm{jet80}}{\lumi^\mathrm{jet80}} & \text{if } \pt > 105\GeV
  \end{cases}
\end{equation}
where $N_\text{jets}^\mathrm{jet\{40, 60, 80\}}$ are the numbers of accepted jets recorded by the ak4PFjet\{40,60,80\} triggers, and $\lumi^\mathrm{jet{\{40,60,80\}}}$ are their luminosities, 0.96, 4.4, and 27.4\pbinv, respectively. Now, the total efficiency $\epsilon$ is consistent with unity, given the use of the full-efficiency \pt windows.
The merging and the resulting detector-level cross section, differential in \pt, are shown in Fig.~\ref{merging}.

\begin{figure}[th!]
 \includegraphics[width=\textwidth]{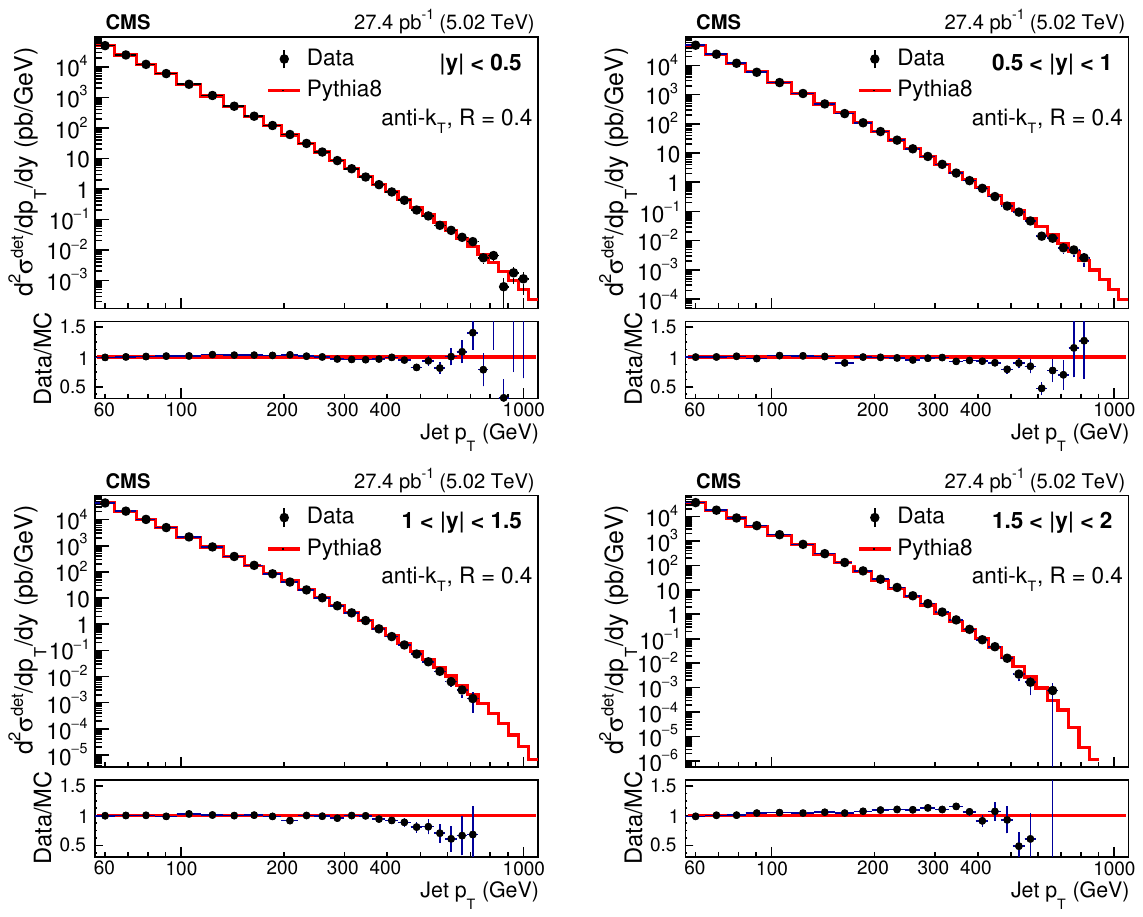}
\caption{Detector-level inclusive jet cross section, differential in \pt, for the four rapidity bins, for the data (points) and the \PYTHIA8 prediction (line) normalized to the total cross section of the data. The lower panels show the ratio of the two distributions. The error bars show the statistical uncertainties in the data.\label{detVsPythia}}
\end{figure}

The detector-level measured inclusive jet cross sections, separately for each of the four rapidity bins, 
can be seen in Fig.~\ref{detVsPythia}. Overlaid is the \PYTHIA8 distribution, 
normalized to the respective detector event yields for shape comparison. Being an LO generator,
\PYTHIA8 is not expected to fully describe the absolute yields and shapes of the observed measured spectra.
Nevertheless, there is a good agreement between the \PYTHIA8 prediction and the observed data in the shape of the detector-level \pt distribution of the cross section in all $\abs{y}$ bins, over several orders of cross section magnitude, as the ratios between data and simulation of Fig.~\ref{detVsPythia} show. For an appropriate comparison with the SM predictions, the detector-level 
spectra need to be corrected to the particle-level spectra through the unfolding procedure (Section~\ref{secUnf}).
The comparisons will be performed against the fixed-order pQCD predictions, corrected for the NP effects presented in the next section.

\section{Perturbative QCD predictions and estimation of NP effects\label{secQCD}}

The next-to-LO (NLO) pQCD prediction for the inclusive jet cross section at 5.02\TeV is determined using the \textsc{NLOJet++} program~\cite{Nagy:2001fj,Nagy:2003tz} (version 4.1.3), whereas the NNLO pQCD prediction is determined with the \textsc{nnlojet}~\cite{NNLOJET,NNLOQCD,infrared,AAPLfast} package (revision 5918), both run within the \textsc{fastNLO} framework~\cite{Britzger:2012bs} (version 2.5). 
The prediction depends on the PDF set used as well as the choice of renormalization scale ($\mu_\mathrm{R}$) and factorization scale ($\mu_\mathrm{F}$).
In this paper, we present the NLO (NNLO) perturbative predictions for the CT18NLO~\cite{ct18} (CT18NNLO~\cite{ct18}, NNPDF3.1NNLO~\cite{nnpdf31})
PDF sets, accessible through the \textsc{lhapdf6} package~\cite{lhapdf6}.
The predictions are computed using one of two scale options: either $\mu = \mu_\mathrm{R} = \mu_\mathrm{F} = \HT$, where \HT is the 
scalar sum of the \pt values of all partons in each event
(an event-based quantity), 
or $\mu = \mu_\mathrm{R} = \mu_\mathrm{F} = \pt$, where \pt is that of each jet in the event (a jet-based quantity). 
All fixed-order pQCD predictions are binned in jet \pt and are subsequently interpolated with a cubic spline,
which allows for a more accurate application of the NP corrections and the subsequent 
rebinning of the cross section distributions through integration. 

\begin{figure}[th!]
\centering
\includegraphics[width=0.6\textwidth]{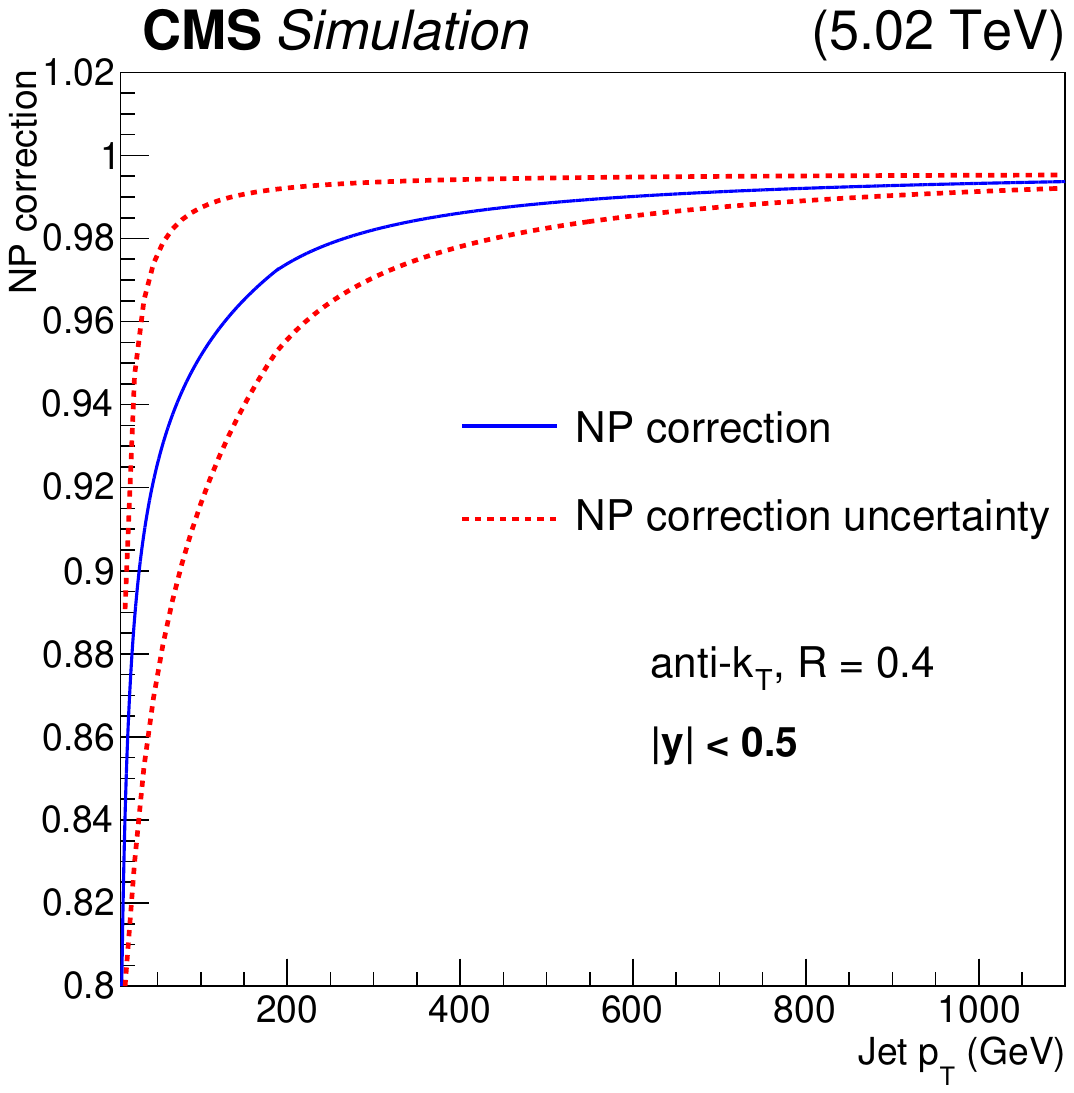}
\caption{Nonperturbative correction to the fixed-order QCD calculation of inclusive jet cross section, as a function of jet \pt, for the $\abs{y} < 0.5$ rapidity bin. Dashed lines show the prediction of corrections using \HERWIG7 (lower line) and \PYTHIA8 (upper line). The central solid line shows the average NP correction used in this analysis, with an uncertainty defined by the extreme predictions. The NP corrections are similar in shape and value for the other rapidity bins.\label{NP}}
\end{figure}

This pQCD prediction has to be corrected for the NP effects of multiparton interactions and hadronization, 
which are estimated using both \PYTHIA8 and \HERWIG7. This estimation is achieved 
by running the generators with and without the NP effects and dividing the corresponding spectra. 
These runs do not include the detector simulation. Because the hadronization and multiparton interaction modeling of the two 
generators differ, we do not expect them to predict the same NP effects. Indeed, the \HERWIG7-derived 
corrections are larger than those of \PYTHIA8. In this analysis, we correct the perturbative spectra
by multiplying them with the average of the two corrections and we use half the difference as a 
systematic uncertainty. Figure~\ref{NP} shows the NP correction as a function of jet \pt, for $\abs{y} < 0.5$, along with 
the systematic uncertainty. Similar distributions are extracted for all rapidity bins.

\section{Unfolding measurement to particle-level spectra\label{secUnf}}

The unfolding method used in this analysis is based on the least-squares minimization, implemented with the \textsc{TUnfold}
\cite{tunfold} package in each rapidity window. The quantity that is minimized is
\begin{equation}
	{L} = (\mathbf{u} - \mathbf{A} {\mathbf{x}} )^{\intercal} {\mathbf{V}}^{-1}(\mathbf{u} - \mathbf{A} \mathbf{x}),
\end{equation}
where $\mathbf{u}$ is the vector of the observed detector-level spectrum (with dimension equal to the number of bins at detector level),
with background removed as described later,
$\mathbf{x}$ is the vector of the unfolded particle-level spectrum (with dimension equal to the number of bins at unfolded level),
$\mathbf{A}$ is the probability matrix, and $\mathbf{V}$ is the data covariance matrix.
The probability matrix is the result of row-wise normalization of the response matrix (described below)
so that each 2D bin content equals the probability that a jet in that particle-level \pt bin is detected in the respective detector-level \pt bin.
The inverse of the data covariance matrix is included to take into account the differing accuracy 
of the elements of the data vector. No explicit regularization or
area constraint is used because they could bias the results. As a form of implicit regularization, the particle-level
binning is coarser than the detector-level binning. This helps to moderate the effect of bin migrations and detector-level 
statistical fluctuations that can cause unphysical behavior in the unfolded result.
The detector-level and particle-level \pt binning for each $\abs{y}$ bin
is shown in Table \ref{tabBins}.

\begin{table}[th!]
\caption{The edges of the \pt bins for the detector-level spectra (all rapidities) and for the particle-level spectra (per rapidity bin). \label{tabBins}}
\center
\begin{tabular}{ c  c  c }
	& $\abs{y}$ binning & \pt binning (\GeVns)\\
\hline
 & & \\
detector-level & all & $\{$56, 64, 74, 84, 97, 114, 133, 153, 174, 196, 220, 245,\\
spectra & rapidities & 272, 300, 330, 362, 395, 430, 468, 507, 548, 592, 638, \\
 & & 686, 737, 790, 846, 905, 967, 1032, 1101$\}$\\
 & & \\
 & 0.0--0.5 & $\{$64, 84, 114, 153, 196, 245, 300, 395, 507, 638, 846$\}$\\
particle-level & 0.5--1.0 & $\{$64, 84, 114, 153, 196, 245, 300, 395, 507, 638, 846$\}$\\
spectra &  1.0--1.5 & $\{$64, 84, 114, 153, 196, 245, 330, 395, 507, 638 $\}$\\
 & 1.5--2.0 & $\{$64, 84, 114, 153, 196, 245, 330, 395, 507$\}$\\      
\end{tabular}
\end{table}

\begin{figure}[th]
  \centering
  \includegraphics[width=0.48\textwidth]{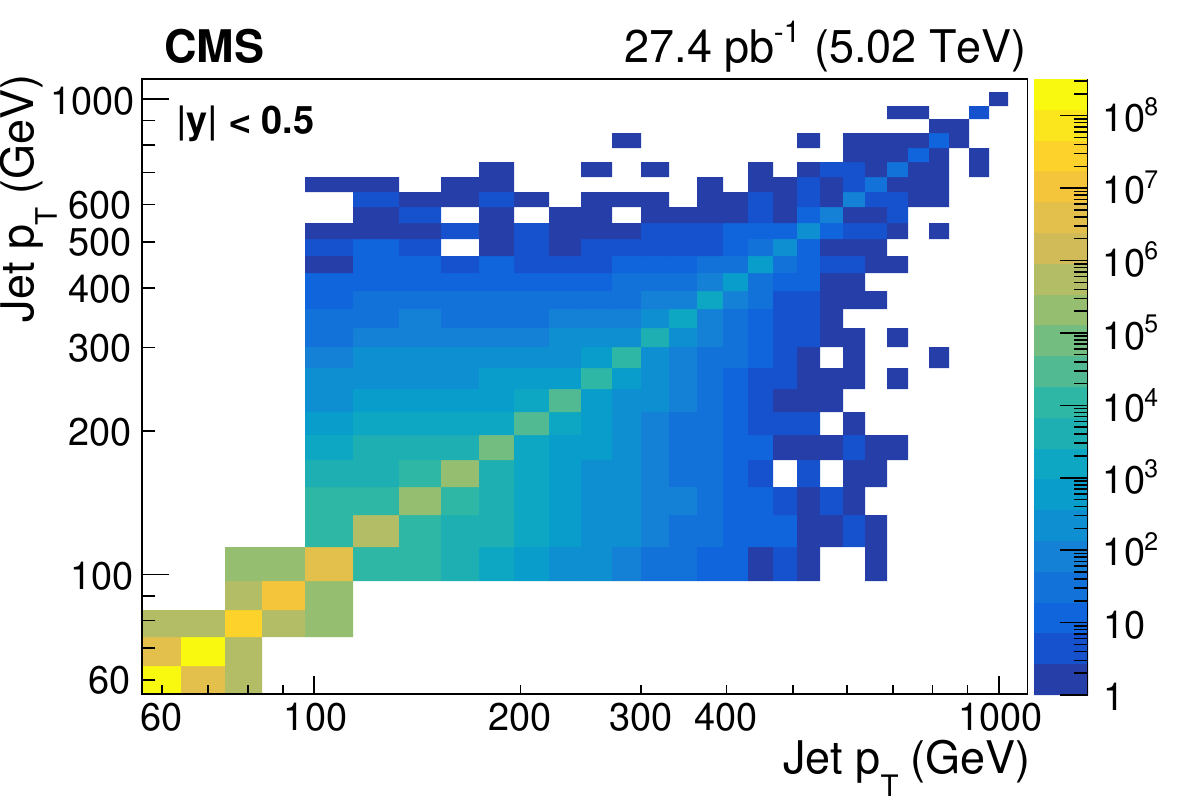}
  \includegraphics[width=0.48\textwidth]{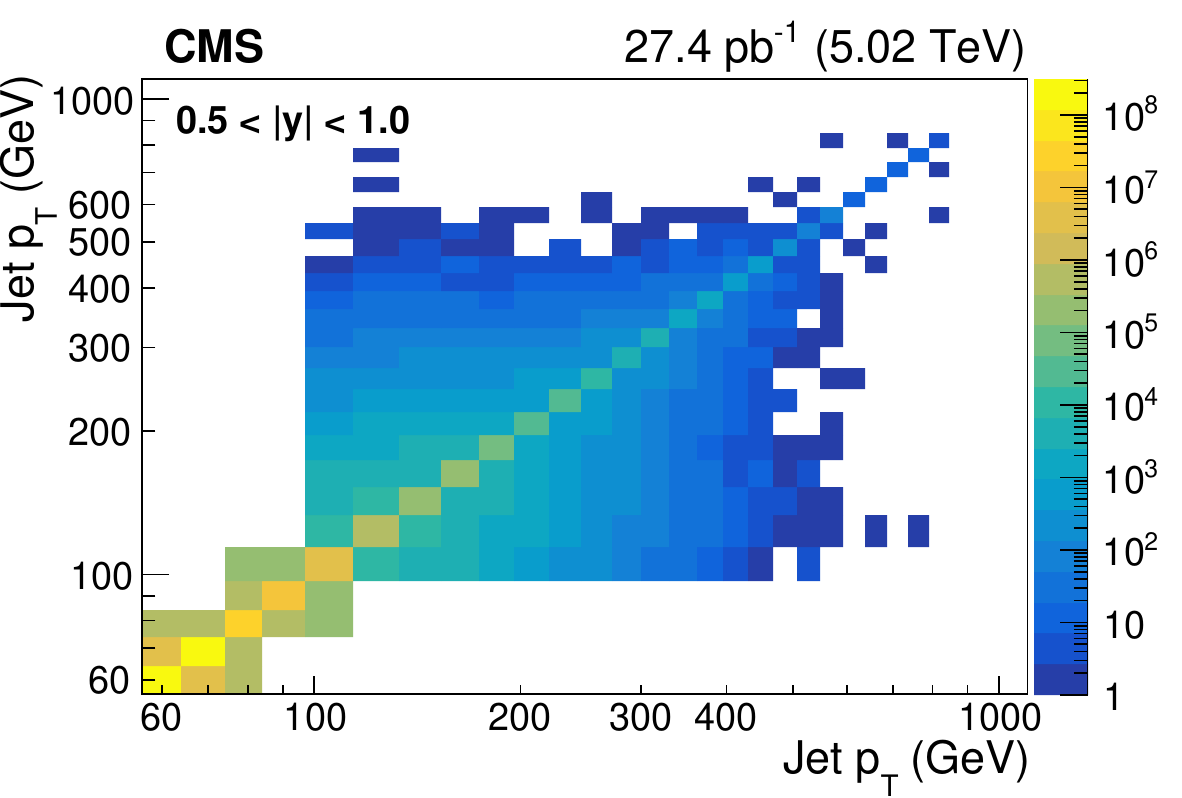}
  \includegraphics[width=0.48\textwidth]{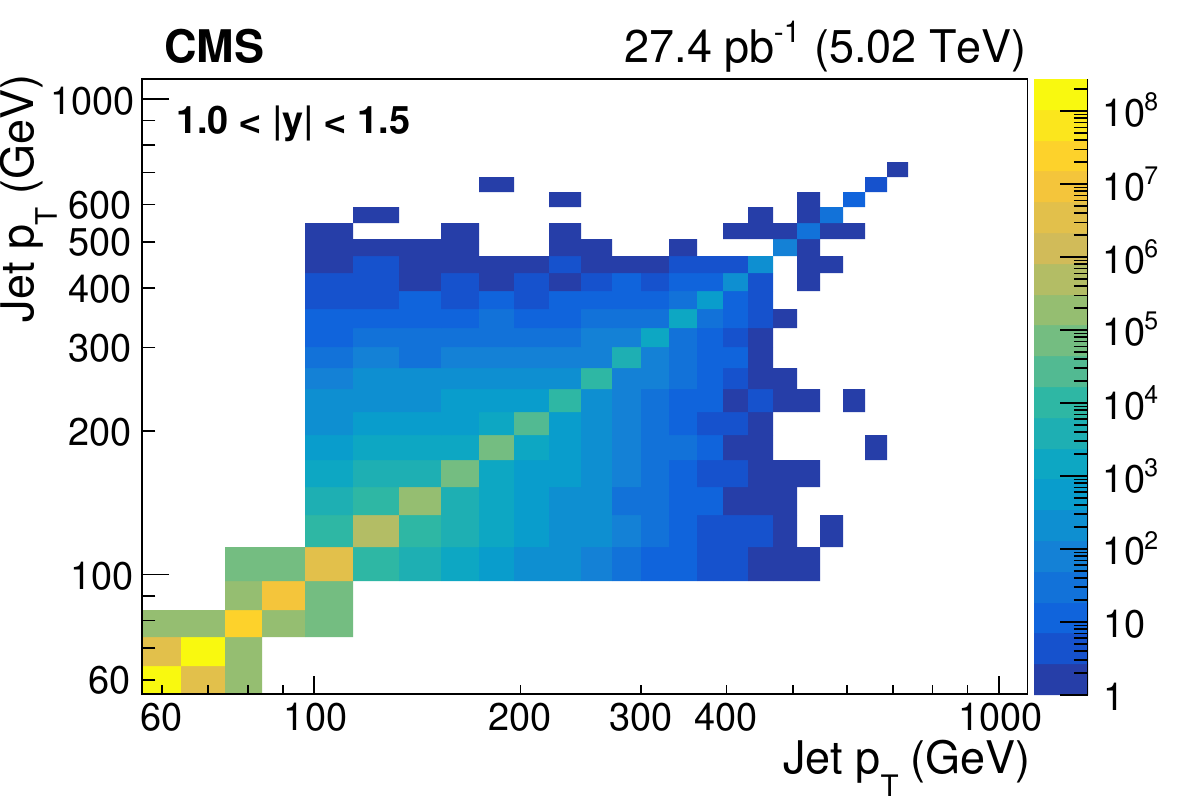}
  \includegraphics[width=0.48\textwidth]{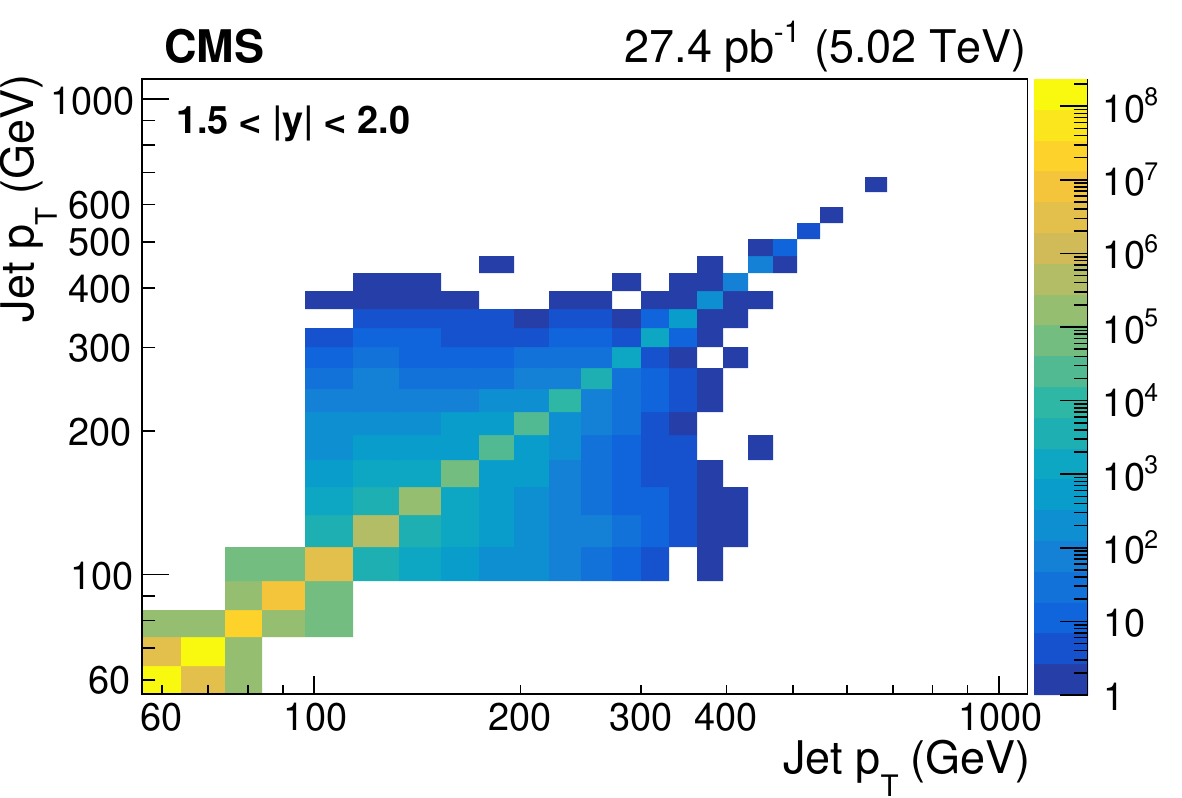}
 \caption{The covariance matrices of the observed detector-level jet \pt for the four rapidity bins. The color scale reports the product of the effective number of jets in the respective \pt bin combinations.\label{CM}}
\end{figure}

The covariance matrices are necessary to properly determine the statistical uncertainty of the measurement 
where multiple jets can be detected per collision event. If $\mathbf{q}$ is the column vector resulting from 
the one-dimensional histogram of measured detector-level jet \pt for an event, then the sum 
of the outer products $\mathbf{q}{\times}\mathbf{q}^\mathrm{T}$ for all detected events constructs a numerical approximation of the two-dimensional covariance matrix. 
Figure~\ref{CM} shows the covariance matrices for the four rapidity bins.
In the unfolding, the entire matrices are used, whereas in the presentation of detector-level spectra (as in Fig.~\ref{detVsPythia}), the square roots of the diagonal elements of the covariance matrices are used to present the statistical uncertainty.

The response matrix is a two-dimensional
histogram filled with the ($\pt^\text{det}$,$\pt^\text{part}$) pairs of detector-level jet \pt and its corresponding 
particle-level jet \pt.
In this analysis we fill this histogram using pseudo-experiments, where we generate random 
\pt values using the expected smooth theoretical spectra (smooth NLO splines multiplied by smooth 
NP corrections, shown in Fig.~\ref{NP} for $\abs{y}<0.5$), to obtain $\pt^\text{part}$, and subsequently
smear them using the JER, to obtain the corresponding $\pt^\text{det}$. 

\begin{figure}[th]
  \centering
  \includegraphics[width=0.48\textwidth]{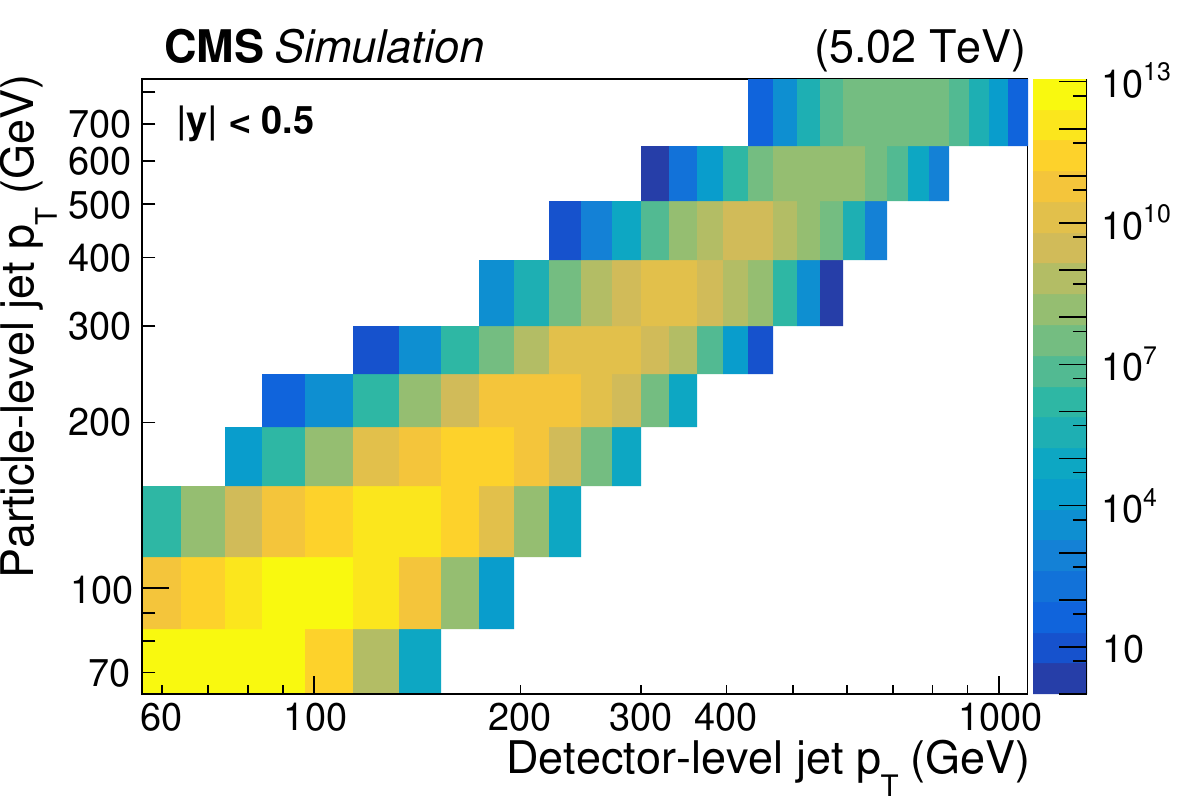}
  \includegraphics[width=0.48\textwidth]{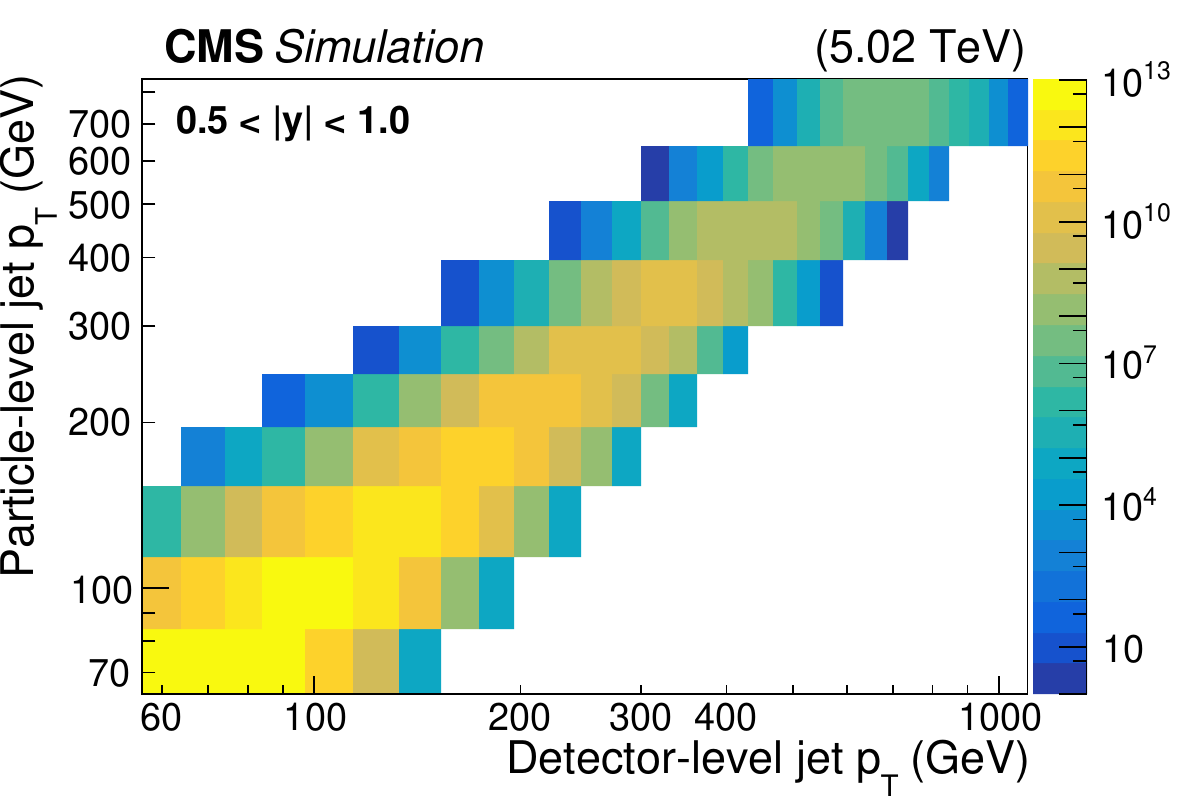}
  \includegraphics[width=0.48\textwidth]{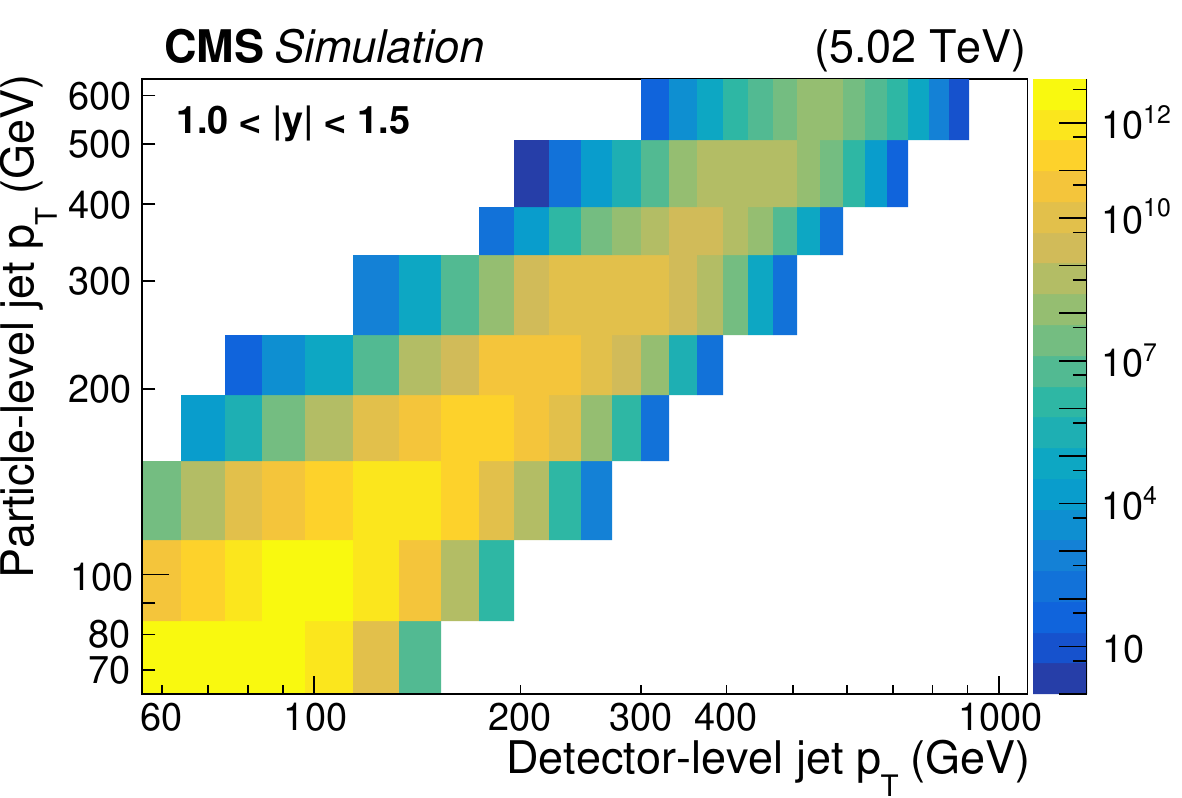}
  \includegraphics[width=0.48\textwidth]{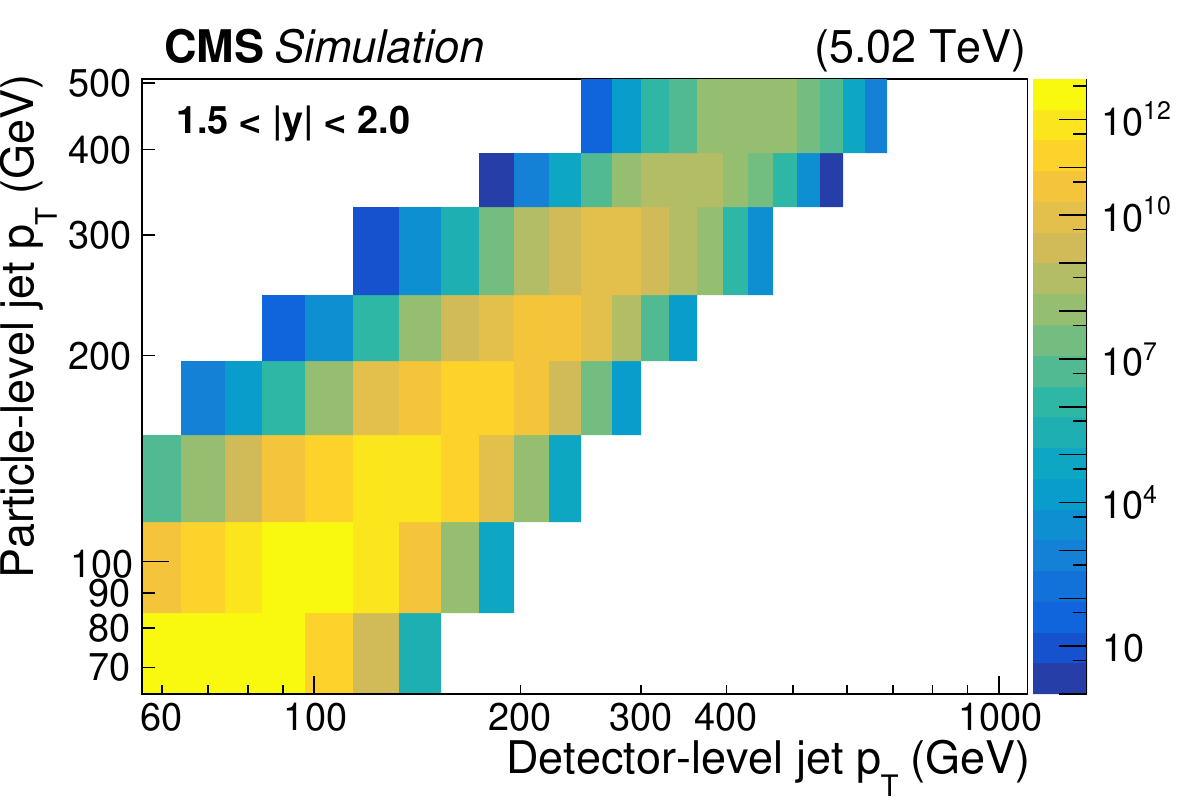}
 \caption{The response matrices for the four rapidity bins. Each 2D-histogram bin contains the number of pseudo-experiment jets that are generated in the particular particle-level \pt bin and that are reconstructed in the corresponding detector-level \pt bin.\label{RM}}
\end{figure}

Figure~\ref{RM} 
shows the response matrices for the four rapidity bins. The range of the detector- and particle-level 
axes corresponds to the detector-level binning and particle-level binning, respectively.
During the filling of the response matrices, we naturally have $\pt^\text{part}$ values within the particle-level
range that give $\pt^\text{det}$ outside the detector-level range. The fraction of these jets gives 
an estimate of the number of the jets missed due to the finite detector resolution. At the same time, 
some $\pt^\text{part}$ values that are outside the particle-level range correspond to $\pt^\text{det}$ values within range. 
The fraction of these jets provides an estimate of background, \ie, jets that migrated to the measured spectrum
from a different particle-level jet kinematic region.
We select the particle-level binning to start at a higher \pt value than the detector-level 
binning (64 vs. 56\GeV) in order to reduce the amount of missed jets 
(they become less than 0.1\% of the jet content of the first \pt bin of the unfolded spectrum and practically 
zero after the third bin).

We subtract the fraction of the background, as estimated above, from the detector spectra of Fig.~\ref{detVsPythia} in order to construct the final detector spectra. We then unfold using the response matrices of Fig.~\ref{RM} and covariance matrices of Fig.~\ref{CM}. Finally, we add the fraction of missed jets to the unfolded spectra to obtain the final particle-level double-differential inclusive jet cross sections to be compared with theoretical predictions. This procedure is repeated for all $\abs{y}$ bins.

Closure tests were performed, where detector-level jet $p_T$ spectra, sampled from the $\pt^\text{det}$ pseudo-experiment distributions, are unfolded and they are compared to the corresponding $\pt^\text{part}$ distributions.  The average agreement over $p_T$ and $y$ bins is $\approx 99.8\%$.
  
Smoothness tests using the tool of Ref.~\cite{patrick} were 
performed on the unfolded results. Smooth representation of the unfolded cross section \pt 
distributions with $\chi^2/\mathrm{dof}<1$ is achieved with 3rd and 4th degree Chebyshev polynomials. 

\section{Systematic uncertainties}

\subsection{Experimental systematic uncertainties}

The main sources of experimental systematic uncertainty are the 
uncertainty in (i) the JES; (ii) the JER; and (iii) the measured luminosity.
Minor sources are the uncertainties in jet identification and the trigger efficiency.
All sources are considered uncorrelated and the total experimental systematic 
uncertainty is derived by summing in quadrature the individual uncertainties.

To estimate the uncertainty in the cross section caused by the uncertainty in the JES, we shift the
\pt of all detector-level jets up and down by one standard deviation of the JES uncertainty,
determined from a number of different sources of jet energy uncertainty, as described in Ref.~\cite{Khachatryan:2016kdb}.
The JES uncertainty in jet \pt is about 1.5\% for \pt of 50\GeV and reduces to about 0.6\% for \pt above 300\GeV. 
For each \pt shift, we repeat the cross section analysis. This includes the merging of the data sets and using the shifted \pt values to determine if the jet passed the threshold at which the trigger is fully efficient. The new detector-level spectra are unfolded with the default response matrices, and the variations of the resulting measured cross sections per \pt and $\abs{y}$ bin are recorded as the systematic uncertainties. 

The JER uncertainty, determined as described in Section~\ref{secJetId}, affects the response matrices used in the unfolding. To estimate 
the effect, alternative response matrices are obtained for each rapidity bin, by shifting the JER up and down according to the uncertainty in the JER scale factor (10\%). We then repeat the unfolding 
for these two response matrices and record the systematic effect on the unfolded
cross sections in all \pt and $\abs{y}$ bins.  The use of JER produced from double-sided Crystal Ball functions with the addition of extra tails
leads to unfolded results shifted downwards on average by $-2.2$\%, consistently across jet \pt and $\abs{y}$ bins.  We increase in quadrature the lower JER systematic uncertainty by this amount. 
Both the JER and JES systematic uncertainties, as functions of \pt, are smoothed by fitting with smooth functions, in order to maximize the value of this data in future PDF and \alpS fits.

The integrated luminosity uncertainty of 2.3\%~\cite{lum16,lum17} shifts the measured cross sections identically in all \pt and $\abs{y}$ bins. Similarly, the jet identification uncertainty introduces a flat 1\% effect~\cite{CMS_7_3} 
on the measured cross section in \pt and $\abs{y}$ bins. Finally, the variation of 
the thresholds for which each trigger is considered to be fully efficient, leads to a 0.3--0.5\% effect on the corresponding threshold \pt bins.

\begin{figure}[ht!]
\centering
\includegraphics[width=\textwidth]{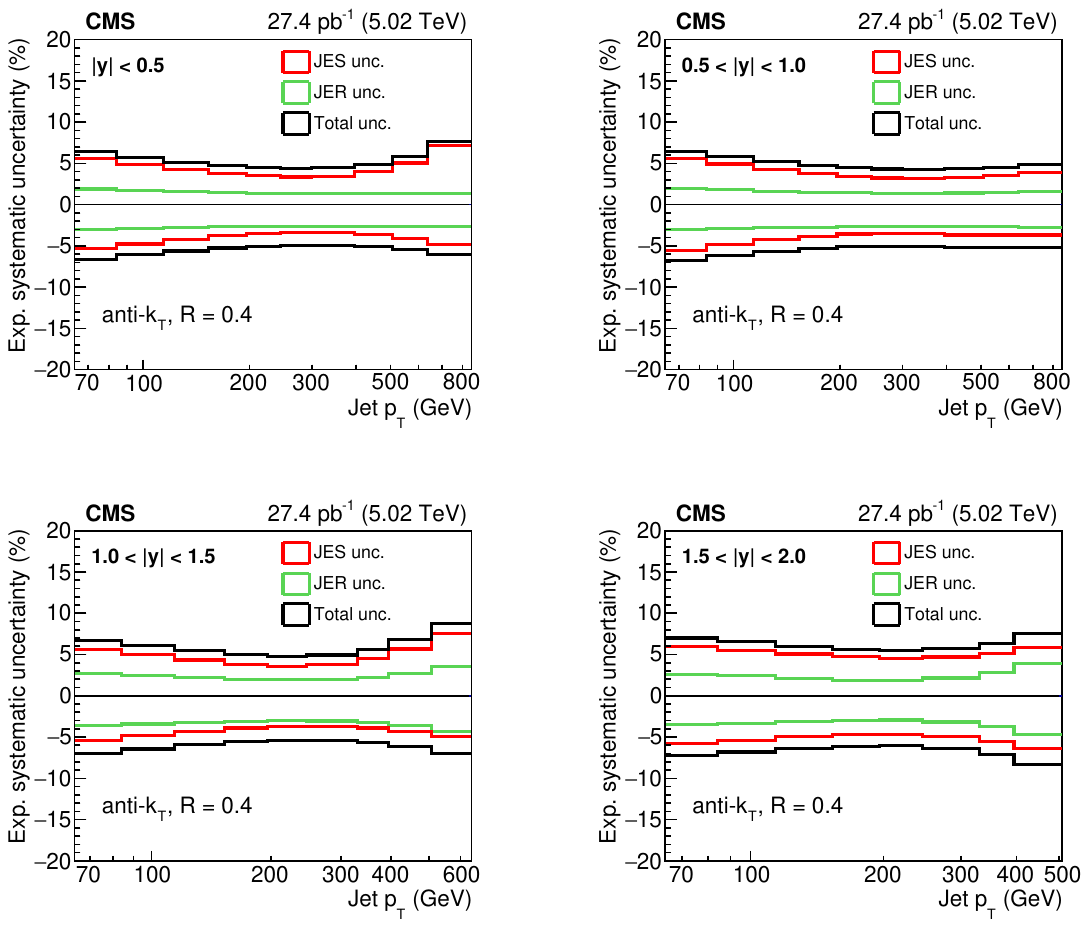}
\caption{The JES, JER, and total systematic uncertainties in unfolded cross sections as functions of transverse momentum and rapidity. The total systematic uncertainty includes also the luminosity, jet identification and trigger efficiency uncertainties.\label{expSyst}}
\end{figure}
There is a negligible effect on the experimental measurement resulting from the
uncertainty in the NP correction, because its effect on the response matrices is extremely small.
This is because this analysis is not sensitive
to very small variations of the prior distribution in the unfolding process.

Figure~\ref{expSyst} shows the effect of JES and JER uncertainties in the unfolded cross sections as 
functions of \pt and $\abs{y}$, as well as the total systematic effect that includes all uncertainties 
summed in quadrature. The uncertainty in JES is the largest source of experimental systematic uncertainty. The total effect 
of the systematic uncertainties in the cross section measurements varies from 4\% to 9\% depending on the \pt and rapidity bin.
The systematic uncertainties of Fig.~\ref{expSyst} originally demonstrated a statistical variation from bin to bin,
which was smoothed out by fitting the \pt-dependence.

\subsection{Theoretical systematic uncertainties}

The main sources of theoretical systematic uncertainties are: (i) the choice of 
$\mu_\mathrm{R}$ and $\mu_\mathrm{F}$ as well as their relationship; (ii) the PDF uncertainty; 
and (iii) the uncertainty in the NP corrections to the fixed-order
calculation. A minor source is the statistical uncertainty from the MC integration. 
 
In this analysis, for the central cross section prediction, we take 
$\mu_\mathrm{F} = \mu_\mathrm{R} = \mu $ and investigate separately
the cases where $\mu$ is equal to either the \pt of each generated jet or \HT. For each case, the maximum
and the minimum variations of the cross section are recorded for the combinations
($\mu_\mathrm{R}$, $\mu_\mathrm{F}$) = $\{$($\mu$, $\mu/2$), ($\mu/2$, $\mu$), ($\mu/2$, $\mu/2$), ($\mu$, 2$\mu$), (2$\mu$, $\mu$), (2$\mu$, 2$\mu$)$\}$. 
The effect depends on the choice of the default scales and is generally larger at higher jet \pt. 
The same is true for the PDF systematic uncertainty, which is related to the PDF fit that leads to a 
particular PDF extraction~\cite{ct18,nnpdf31}. On the other hand, the NP correction uncertainty 
affects mostly the jets with lower \pt, as shown in Fig.~\ref{NP} for $\abs{y}<0.5$.

The above sources are considered uncorrelated, and the total theoretical systematic uncertainty 
is determined by summing in quadrature the individual uncertainties.
 
The distributions of the theoretical systematic uncertainty in \pt and $\abs{y}$ depend on the fixed-order calculation 
in pQCD, and the choice of PDF, $\mu_\mathrm{R}$, and $\mu_\mathrm{F}$; they will be presented in the next section. 

\section{Results and comparisons with theoretical predictions}

\subsection{Results}

\begin{figure}[htb!]
\centering
\includegraphics[width=0.9\textwidth]{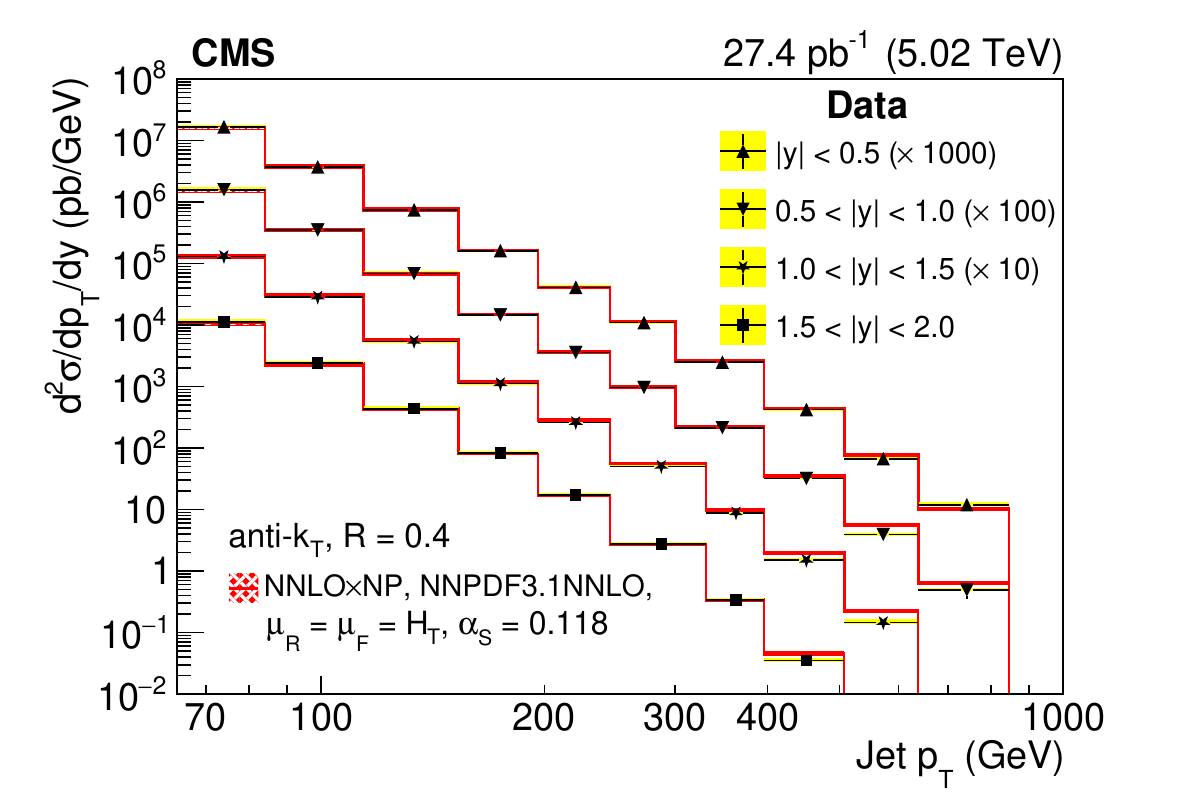}
\caption{The unfolded measured particle-level inclusive jet cross sections as functions of jet \pt in the four rapidity bins (markers), compared to the NNLO perturbative QCD prediction (red histogram), using the NNPDF3.1NNLO PDF set, with $\mu_\mathrm{R}=\mu_\mathrm{F}=\HT$, and corrected for the NP effects. The yellow (red) band shows the experimental (theoretical) systematic uncertainty. Statistical uncertainties are included but are barely visible.\label{unfoldedVsTheoryWithSystematics}}
\end{figure}

Figure~\ref{unfoldedVsTheoryWithSystematics} shows the unfolded
double-differential inclusive particle-level jet cross sections, across 7 orders of magnitude, as functions of \pt for 
the four rapidity bins. The measurements are compared with the theoretical predictions from the NNLO pQCD calculation, with $\mu = \HT$, using the NNPDF3.1NNLO PDF set, and corrected for NP effects. 
To get these theoretical predictions, we multiply the \pt-smooth spline interpolation of the pQCD prediction with the smooth NP corrections, shown in Fig.~\ref{NP} for $\abs{y}<0.5$, and for each bin we integrate this distribution and divide by the bin width.
The vertical error bars correspond to the statistical uncertainty in the measurement and the yellow band around the experimental measurement represents the total systematic uncertainty. This systematic uncertainty varies from 4 to 9\%, as shown in Fig.~\ref{expSyst}, whereas the statistical uncertainty
ranges from 0.2\% to 27\% and becomes dominant above jet \pt of 500\GeV for the first three rapidity bins and above 400\GeV for the last one.
The red band around the theory prediction shows the total theoretical systematic uncertainty. 

\subsection{Comparisons with theoretical predictions}

The differences between experimental measurements and theoretical predictions are revealed
in the ratios of the respective cross sections in all \pt bins and $\abs{y}$ bins.
Because the sources of experimental and theoretical uncertainties are
independent, the respective systematic uncertainties are decoupled.
For the purpose of better visual comparisons, we center the experimental
uncertainties around the ratio points and the theoretical uncertainties
around the line at one.
The experimental cross sections and their uncertainties, as well as the uncertainties in the theory,
are all divided by a constant theoretical cross section. For these ratios, the uncertainties appear only in the numerators.
\begin{figure}[thb!]
\includegraphics[width=0.95\textwidth]{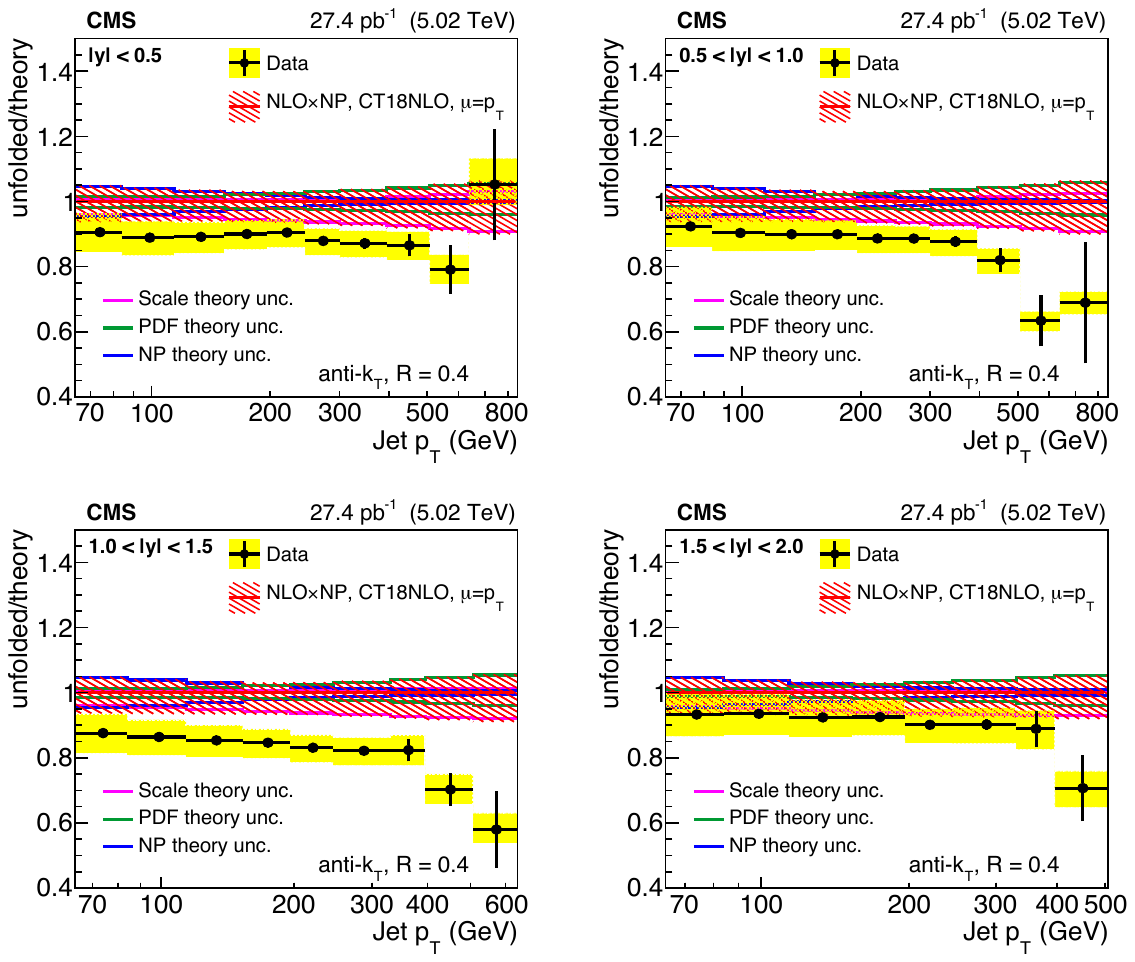}
\caption{Ratios (points) of the unfolded measured cross sections to the NLO theoretical predictions, using the CT18NLO PDF set, with $\mu = \pt$. The vertical error bars show the statistical experimental uncertainty, the yellow band shows the systematic experimental uncertainty, the hashed red band shows the total theoretical uncertainty, and the individual sources of theoretical uncertainty are shown with colored lines.\label{4}}
\end{figure}
\begin{figure}[thb!]
\includegraphics[width=0.95\textwidth]{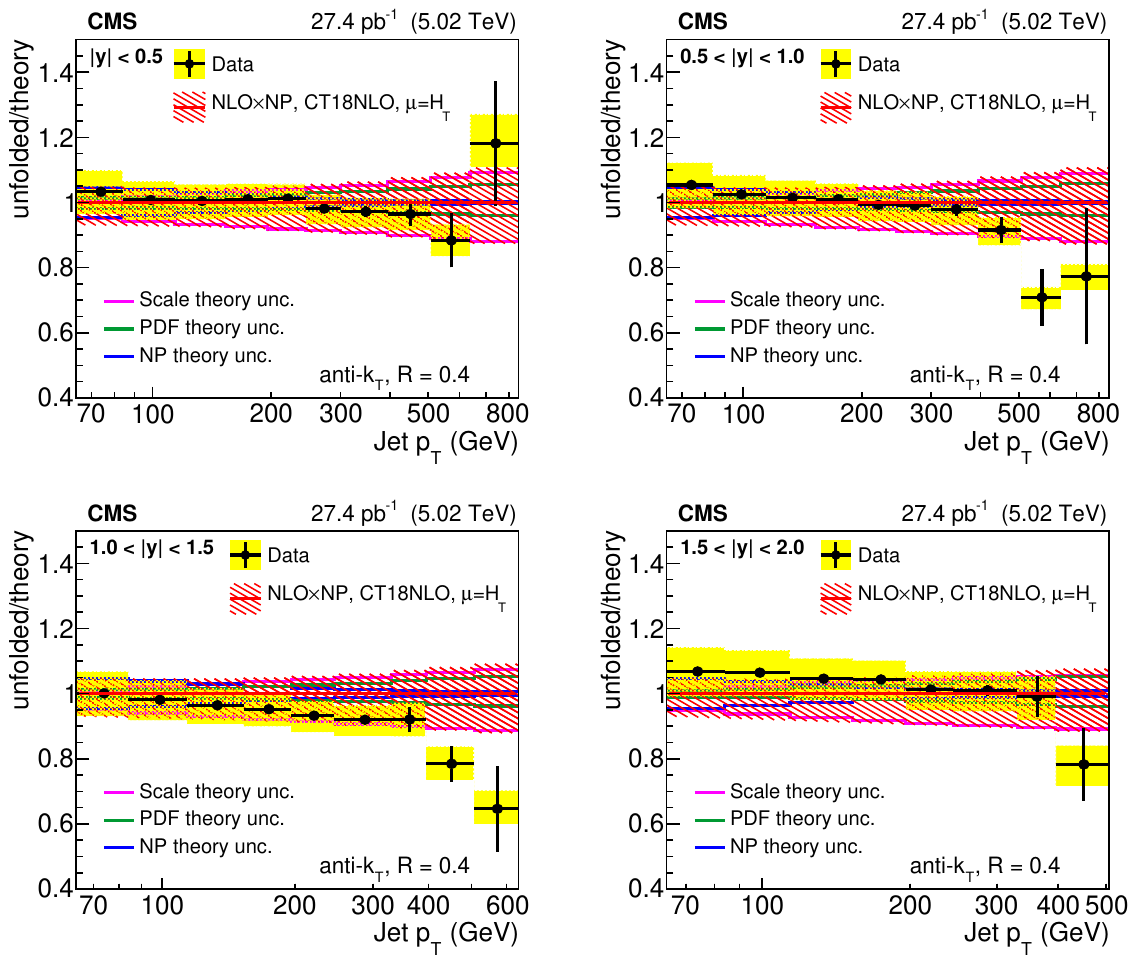}
\caption{Ratios (points) of the unfolded measured cross sections to the NLO theoretical predictions, using the CT18NLO PDF set, with $\mu= \HT$. The vertical error bars show the statistical experimental uncertainty, the yellow band shows the systematic experimental uncertainty, the hashed red band shows the total theoretical uncertainty, and the individual sources of theoretical uncertainty are shown with colored lines.\label{104}}
\end{figure}

Figure~\ref{4} shows the ratios of the unfolded inclusive jet cross sections to the NLO theoretical predictions, using the CT18NLO PDF set, with $\mu_\mathrm{R} = \mu_\mathrm{F} = \pt$, as functions of \pt, in the four $\abs{y}$ bins.
The theoretical predictions are systematically larger compared to the data, mostly for $\abs{y}<1.5$.

Figure~\ref{104} shows the ratios of the unfolded inclusive jet cross sections to the NLO theoretical predictions, using the CT18NLO PDF set, with $\mu_\mathrm{R} = \mu_\mathrm{F} = \HT$, as functions of \pt, in the four $\abs{y}$ bins. We observe a good agreement between theory and experiment, 
with data being slightly above the theoretical prediction for $\abs{y} > 1.5$.
We also observe larger theoretical systematic uncertainty in the predictions with $\mu=\HT$ compared to $\mu=\pt$, coming from the larger choice of scale magnifying the uncertainties, especially at higher \pt.

\begin{figure}[htb!]
\includegraphics[width=0.95\textwidth]{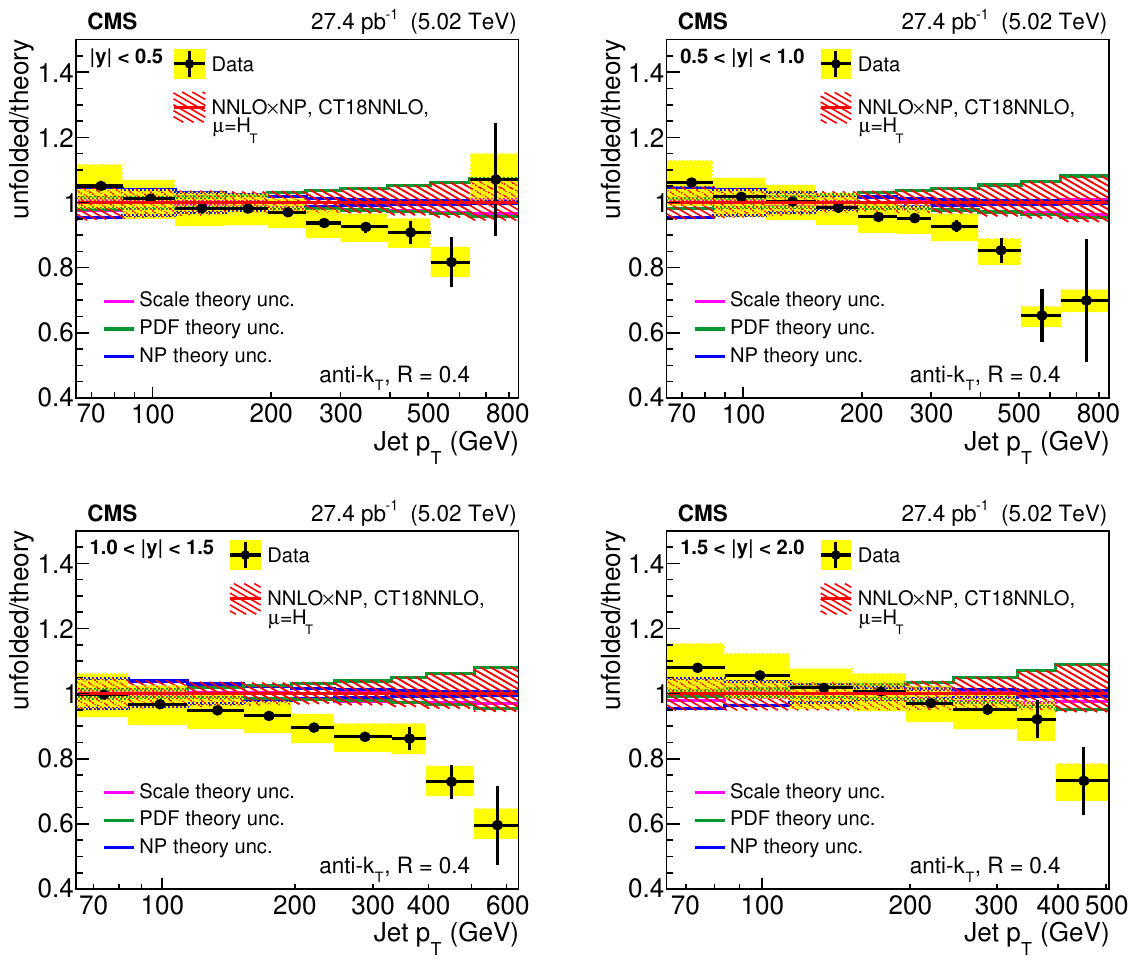}
\caption{Ratios (points) of the unfolded measured cross sections to the NNLO theoretical predictions, using the CT18NNLO PDF set, with $\mu=\HT$. The vertical error bars show the statistical experimental uncertainty, the yellow band shows the systematic experimental uncertainty, the hashed red band shows the total theoretical uncertainty, and the individual sources of theoretical uncertainty are shown with colored lines.\label{1104}}
\end{figure}

Figure~\ref{1104} shows the ratios of the unfolded inclusive jet cross sections to the NNLO theoretical predictions, using the CT18NNLO PDF set, with $\mu_\mathrm{R} = \mu_\mathrm{F} = \HT$, as functions of \pt, in the four $\abs{y}$ bins.
We observe a ratio which is less uniform across \pt, 
demonstrating an underestimation from theory at lower \pt and an overestimation at higher \pt. 
Also, the theoretical systematic uncertainties at higher \pt are smaller for the NNLO prediction, because of the reduced effects of scale variations.
The fluctuations in the ratio comes from the statistical uncertainty in the MC integration in the NNLO calculation.

\begin{figure}[htb!]
\includegraphics[width=0.95\textwidth]{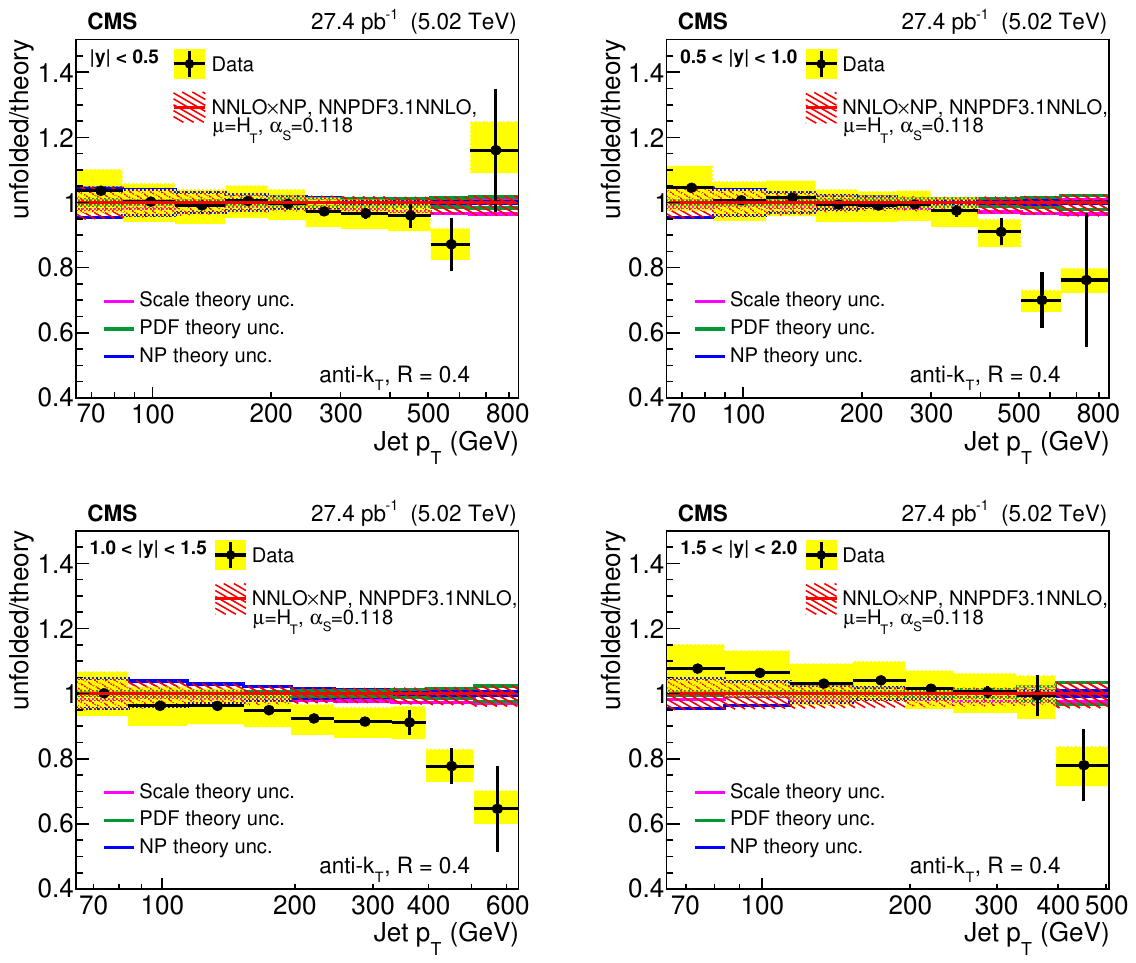}
\caption{Ratios (points) of the unfolded measured cross sections to the NNLO theoretical predictions, using the NNPDF3.1NNLO PDF set, with $\alpS(m_{\Z})=0.118$ and $\mu=\HT$. The vertical error bars show the statistical experimental uncertainty, the yellow band shows the systematic experimental uncertainty, the hashed red band shows the total theoretical uncertainty, and the individual sources of theoretical uncertainty are shown with colored lines.\label{1124}}
\end{figure}

Figure~\ref{1124} shows the ratios of the unfolded inclusive jet cross sections to the NNLO theoretical predictions,
using the NNPDF3.1NNLO ($\alpS(m_{\Z})=0.118$) PDF set, for $\mu_\mathrm{R} = \mu_\mathrm{F} = \HT$, as functions of \pt, in the four $\abs{y}$ bins.
Compared with the CT18NNLO PDF set, we observe a flatter ratio and a smaller theoretical systematic uncertainty associated with the NNPDF3.1NNLO set. This particular theoretical prediction agrees well with the experimental results for jet $\pt<400\GeV$, both in uniformity of the ratios across \pt and their proximity to unity.

\begin{figure}[htp!]
\includegraphics[width=0.95\textwidth]{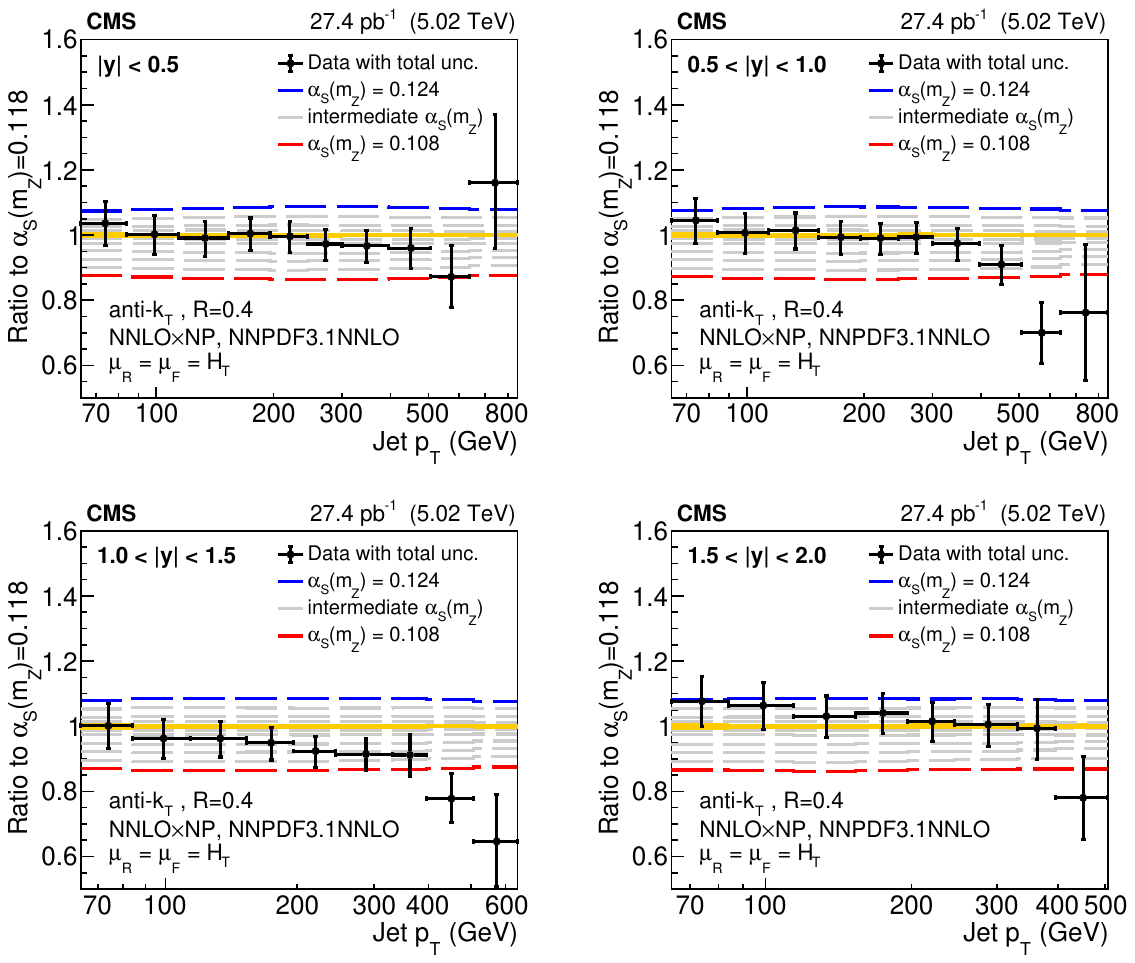}
\caption{The effect of $\alpS(m_{\Z})$ variation. The NNLO theoretical cross section predictions 
using the NNPDF3.1NNLO PDF with $\mu=\HT$,
calculated for different choices of \alpS (0.108, 0.110, 0.112, 0.114, 0.116, 0.117, 0.118, 0.119, 0.120, 0.122, 
and 0.124), are divided by the benchmark NNLO prediction for $\alpS = 0.118$ and the same choice of PDF set, $\mu_\mathrm{R}$, and $\mu_\mathrm{F}$. Also shown is the experimental unfolded measurement divided by the same benchmark prediction. The width of the unity line corresponds to the statistical uncertainty from the MC integration for the determination of the NNLO prediction. The error bars on the unfolded data correspond to the total experimental statistical and systematic uncertainty added in quadrature.\label{multiAs}}
\end{figure}

Figure~\ref{multiAs} shows the effect of $\alpS(m_{\Z})$ variation on the NNLO theoretical predictions and the comparison with 
experimental results, for NNPDF3.1NNLO and $\mu=\HT$. The values of \alpS range from 0.108 to 0.124.
In this plot, the unfolded experimental measurement and 
all theoretical predictions for different \alpS choices are divided by 
a benchmark NNLO theoretical prediction for $\alpS = 0.118$ and the same choice of PDFs and scales.

From these results we conclude that the NLO calculation using the CT18NLO PDF set with $\mu = \HT$ agrees better with the experimental measurement than the same calculation with $\mu = \pt$. The latter calculation gives a higher cross section across all \pt bins for all $\abs{y}$ bins, maintaining
the same shape. On the other hand, the NNLO prediction changes the shape of the ratio for the CT18NNLO PDF set, underestimating the cross section at lower \pt bins and overestimating it at higher
\pt bins, across all $\abs{y}$ bins. The scale systematic uncertainties are significantly decreased at high \pt
in NNLO calculations and the NNPDF3.1NNLO PDF set additionally provides small PDF-related systematic uncertainties.
These observations provide valuable input to the understanding of jet production, demonstrating how well the data are described by QCD calculations, using various PDF sets, and choices of $\mu_\mathrm{R}$ and $\mu_\mathrm{F}$.

\clearpage
\section{Summary}

The double-differential inclusive jet cross section
in proton-proton collisions at 5.02\TeV was measured
in the rapidity interval $\abs{y} < 2$, and for the transverse momentum range $0.06 < \pt < 1\TeV$. The achieved experimental systematic uncertainty is about 5\% across most \pt ranges for all $\abs{y}$. 
The next-to-leading order (NLO) perturbative quantum chromodynamics calculations agree better with the observations if the renormalization and factorization scales ($\mu$) equal \HT, the scalar sum of the transverse momentum (\pt) of the partons in each event.
The energy scale systematic uncertainty also increases when the scale is changed from
$\mu=\pt$ of each jet to $\mu=\HT$ in the NLO case.
Changing the order of the perturbative calculation from the NLO to next-to-NLO (NNLO) reduces 
the scale systematic uncertainty at high \pt,
but increases it at low \pt. The effect of changing the scale is not 
very large for the NNLO calculation, and the scale systematic decreases
at low \pt when the scale is changed from $\mu=\pt$ to $\mu=\HT$.
The uncertainty in the predicted cross section due to the parton distribution functions is significantly reduced by choosing the NNPDF3.1NNLO set.  The measurement is consistent with NLO and the more precise NNLO prediction.

\begin{acknowledgments}

We thank Jo\~ao Pires for his work on the NNLO pQCD predictions.

We congratulate our colleagues in the CERN accelerator departments for the excellent performance of the LHC and thank the technical and administrative staffs at CERN and at other CMS institutes for their contributions to the success of the CMS effort. In addition, we gratefully acknowledge the computing centers and personnel of the Worldwide LHC Computing Grid and other centers for delivering so effectively the computing infrastructure essential to our analyses. Finally, we acknowledge the enduring support for the construction and operation of the LHC, the CMS detector, and the supporting computing infrastructure provided by the following funding agencies: SC (Armenia), BMBWF and FWF (Austria); FNRS and FWO (Belgium); CNPq, CAPES, FAPERJ, FAPERGS, and FAPESP (Brazil); MES and BNSF (Bulgaria); CERN; CAS, MoST, and NSFC (China); MINCIENCIAS (Colombia); MSES and CSF (Croatia); RIF (Cyprus); SENESCYT (Ecuador); MoER, ERC PUT and ERDF (Estonia); Academy of Finland, MEC, and HIP (Finland); CEA and CNRS/IN2P3 (France); SRNSF (Georgia); BMBF, DFG, and HGF (Germany); GSRI (Greece); NKFIH (Hungary); DAE and DST (India); IPM (Iran); SFI (Ireland); INFN (Italy); MSIP and NRF (Republic of Korea); MES (Latvia); LAS (Lithuania); MOE and UM (Malaysia); BUAP, CINVESTAV, CONACYT, LNS, SEP, and UASLP-FAI (Mexico); MOS (Montenegro); MBIE (New Zealand); PAEC (Pakistan); MES and NSC (Poland); FCT (Portugal); MESTD (Serbia); MCIN/AEI and PCTI (Spain); MOSTR (Sri Lanka); Swiss Funding Agencies (Switzerland); MST (Taipei); MHESI and NSTDA (Thailand); TUBITAK and TENMAK (Turkey); NASU (Ukraine); STFC (United Kingdom); DOE and NSF (USA).

\hyphenation{Rachada-pisek} Individuals have received support from the Marie-Curie program and the European Research Council and Horizon 2020 Grant, contract Nos.\ 675440, 724704, 752730, 758316, 765710, 824093, and COST Action CA16108 (European Union); the Leventis Foundation; the Alfred P.\ Sloan Foundation; the Alexander von Humboldt Foundation; the Science Committee, project no. 22rl-037 (Armenia); the Belgian Federal Science Policy Office; the Fonds pour la Formation \`a la Recherche dans l'Industrie et dans l'Agriculture (FRIA-Belgium); the Agentschap voor Innovatie door Wetenschap en Technologie (IWT-Belgium); the F.R.S.-FNRS and FWO (Belgium) under the ``Excellence of Science -- EOS" -- be.h project n.\ 30820817; the Beijing Municipal Science \& Technology Commission, No. Z191100007219010 and Fundamental Research Funds for the Central Universities (China); the Ministry of Education, Youth and Sports (MEYS) of the Czech Republic; the Shota Rustaveli National Science Foundation, grant FR-22-985 (Georgia); the Deutsche Forschungsgemeinschaft (DFG), under Germany's Excellence Strategy -- EXC 2121 ``Quantum Universe" -- 390833306, and under project number 400140256 - GRK2497; the Hellenic Foundation for Research and Innovation (HFRI), Project Number 2288 (Greece); the Hungarian Academy of Sciences, the New National Excellence Program - \'UNKP, the NKFIH research grants K 124845, K 124850, K 128713, K 128786, K 129058, K 131991, K 133046, K 138136, K 143460, K 143477, 2020-2.2.1-ED-2021-00181, and TKP2021-NKTA-64 (Hungary); the Council of Science and Industrial Research, India; ICSC -- National Research Center for High Performance Computing, Big Data and Quantum Computing, funded by the EU NexGeneration program (Italy); the Latvian Council of Science; the Ministry of Education and Science, project no. 2022/WK/14, and the National Science Center, contracts Opus 2021/41/B/ST2/01369 and 2021/43/B/ST2/01552 (Poland); the Funda\c{c}\~ao para a Ci\^encia e a Tecnologia, grant CEECIND/01334/2018 (Portugal); the National Priorities Research Program by Qatar National Research Fund; MCIN/AEI/10.13039/501100011033, ERDF ``a way of making Europe", and the Programa Estatal de Fomento de la Investigaci{\'o}n Cient{\'i}fica y T{\'e}cnica de Excelencia Mar\'{\i}a de Maeztu, grant MDM-2017-0765 and Programa Severo Ochoa del Principado de Asturias (Spain); the Chulalongkorn Academic into Its 2nd Century Project Advancement Project, and the National Science, Research and Innovation Fund via the Program Management Unit for Human Resources \& Institutional Development, Research and Innovation, grant B37G660013 (Thailand); the Kavli Foundation; the Nvidia Corporation; the SuperMicro Corporation; the Welch Foundation, contract C-1845; and the Weston Havens Foundation (USA).
\end{acknowledgments}

\bibliography{auto_generated}
\cleardoublepage \appendix\section{The CMS Collaboration \label{app:collab}}\begin{sloppypar}\hyphenpenalty=5000\widowpenalty=500\clubpenalty=5000
\cmsinstitute{Yerevan Physics Institute, Yerevan, Armenia}
{\tolerance=6000
A.~Hayrapetyan, A.~Tumasyan\cmsAuthorMark{1}\cmsorcid{0009-0000-0684-6742}
\par}
\cmsinstitute{Institut f\"{u}r Hochenergiephysik, Vienna, Austria}
{\tolerance=6000
W.~Adam\cmsorcid{0000-0001-9099-4341}, J.W.~Andrejkovic, T.~Bergauer\cmsorcid{0000-0002-5786-0293}, S.~Chatterjee\cmsorcid{0000-0003-2660-0349}, K.~Damanakis\cmsorcid{0000-0001-5389-2872}, M.~Dragicevic\cmsorcid{0000-0003-1967-6783}, A.~Escalante~Del~Valle\cmsorcid{0000-0002-9702-6359}, P.S.~Hussain\cmsorcid{0000-0002-4825-5278}, M.~Jeitler\cmsAuthorMark{2}\cmsorcid{0000-0002-5141-9560}, N.~Krammer\cmsorcid{0000-0002-0548-0985}, L.~Lechner\cmsorcid{0000-0002-3065-1141}, D.~Liko\cmsorcid{0000-0002-3380-473X}, I.~Mikulec\cmsorcid{0000-0003-0385-2746}, J.~Schieck\cmsAuthorMark{2}\cmsorcid{0000-0002-1058-8093}, R.~Sch\"{o}fbeck\cmsorcid{0000-0002-2332-8784}, D.~Schwarz\cmsorcid{0000-0002-3821-7331}, M.~Sonawane\cmsorcid{0000-0003-0510-7010}, S.~Templ\cmsorcid{0000-0003-3137-5692}, W.~Waltenberger\cmsorcid{0000-0002-6215-7228}, C.-E.~Wulz\cmsAuthorMark{2}\cmsorcid{0000-0001-9226-5812}
% ...
\par}
\vskip\cmsinstskip
\dag:~Deceased\\
$^{1}$Also at Yerevan State University, Yerevan, Armenia\\
$^{2}$Also at TU Wien, Vienna, Austria\\
% ...
\end{sloppypar}
\end{document}